\pdfoutput=1
\RequirePackage{ifpdf}
\documentclass[paper,hyper]{JHEP3}
\usepackage{epsfig,subfig}
\usepackage{graphicx}
\usepackage{amssymb}
\usepackage{amsmath}
\usepackage{booktabs,multirow,tabularx}
\usepackage{slashed}
\usepackage{cite}
\usepackage{caption}
\usepackage{float}
\usepackage{placeins}
\usepackage{rotating}
\usepackage{lscape}

\newcommand{\bqa}{\begin{eqnarray}}
\newcommand{\eqa}{\end{eqnarray}}

\def\beq{\begin{equation}}
\def\eeq{\end{equation}}
\def\beqn{\begin{eqnarray}}
\def\eeqn{\end{eqnarray}}
\def\abs#1{\left|#1\right|}

\newcommand\sss{\scriptscriptstyle}

\newcommand\as{\alpha_{\sss S}}

\newcommand{\muF}{\mu_{\sss F}}
\newcommand{\muR}{\mu_{\sss R}}
\newcommand\MadFKS{{\sc\small MadFKS}}
\newcommand\CutTools{{\sc\small CutTools}}
\newcommand\MadLoop{{\sc\small MadLoop}}

\newcommand\aNLO{{\sc\small MadGraph5\_aMC@NLO}}

\newcommand\HWs{{\sc\small HERWIG6}}
\newcommand\HWpp{{\sc\small Herwig++}}
\newcommand\PY{{\sc\small Pythia}}

\newcommand\PYe{{\sc\small Pythia8}}

\newcommand\UFO{{\sc\small UFO}}

\newcommand\OL{{\sc\small OpenLoops}}

\newcommand\prompt{{\tt MG5\_aMC>}}
\newcommand\bb{\bar{b}}
\newcommand\bq{\bar{q}}
\newcommand\bbH{b\bar{b}H}
\newcommand\mb{m_b}
\newcommand\mH{m_H}
\newcommand\yt{y_t}
\newcommand\yb{y_b}
\newcommand{\pt}{p_{\sss T}}
\newcommand{\kt}{k_{\sss T}}
\newcommand{\Ht}{H_{\sss T}}
\newcommand{\mt}{m_{\sss T}}
\newcommand{\ptsyst}{p_{\sss T}^{\rm syst}}
\newcommand{\LQCD}{\Lambda_{\sss\rm QCD}}

\def\bal#1\eal{\begin{align}#1\end{align}}

\newcommand{\abbrev}{}

\newcommand{\mtop}{m_t}
\newcommand{\mbottom}{m_b}

\newcommand{\ybyt}{\ensuremath{y_by_t}}

\newcommand{\lo}{{\abbrev LO}}

\newcommand{\sm}{\rm SM}

\newcommand{\thdm}{{\abbrev 2HDM}}

\newcommand{\fs}[1]{#1{\abbrev FS}}

\newcommand{\bbh}{b\bar bH}

\newcommand{\Qres}{Q_{\text{res}}}
\newcommand{\Qshow}{Q_{\text{sh}}}
\newcommand{\Qshowmax}{\Qshow}

\newcommand{\msbar}{{\abbrev \overline{\text{MS}}}}
\newcommand{\os}{{o.s.}}

\preprint{
 CERN-PH-TH-2014-182\\
 CP3-14-64, LPN14-114\\
 MCNET-14-20, ZU-TH 33/14
 }
\title{Higgs production in association with bottom quarks}
\author{M.~Wiesemann$^a$, R.~Frederix$^b$, S.~Frixione$^b$,
V.~Hirschi$^c$, F.~Maltoni$^d$, P.~Torrielli$^{ae}$\\
 $^a$ Physik-Institut, Universit\"at Z\"urich, Winterthurerstrasse 190, 
8057 Zurich, Switzerland\\
 $^b$ PH Department, TH Unit, CERN, CH-1211 Geneva 23, Switzerland\\
 $^c$ SLAC National Accelerator Laboratory, 2575 Sand Hill Road,\\
 $\phantom{^e}$ Menlo Park, CA 94025-7090 USA\\
 $^d$ CP3, \mbox{Universit\'e Catholique de Louvain}, Chemin du Cyclotron 2, \\
  $\phantom{^e}$ B-1348 Louvain-la-Neuve, Belgium\\
 $^e$ Dipartimento di Fisica, Universit\`a di Torino, Via P.~Giuria~1, 
 I-10125 Torino, Italy\\
}

\abstract{
We study the production of a Higgs boson in association with bottom 
quarks in hadronic collisions, and present phenomenological predictions
relevant to the 13~TeV LHC. Our results are accurate to the next-to-leading
order in QCD, and matched to parton showers through the MC@NLO method;
thus, they are fully differential and based on unweighted events,
which we shower by using both {\sc\small Herwig++} and {\sc\small Pythia8}.
We perform the computation in both the four-flavour and the five-flavour
schemes, whose results we compare extensively at the level of exclusive
observables. In the case of the Higgs transverse momentum, we also consider 
the analytically-resummed cross section up to the NNLO+NNLL accuracy. 
In addition, we analyse at ${\cal O}(\alpha_S^3)$ the effects of the 
interference between the $\bbH$ and gluon-fusion production modes.
}
\keywords{QCD, NLO Computations, Higgs}

\begin{document}

\section{Introduction\label{sec:intro}}
Data collected at the LHC so far fully support the hypothesis that 
the observed resonance with a mass of about 125 GeV is the scalar 
boson predicted by the Brout-Englert-Higgs symmetry breaking 
mechanism~\cite{Englert:1964et,Higgs:1964pj} of $SU(2)_L \times U(1)_Y$, as
implemented in the Standard Model 
(SM)~\cite{Glashow:1961tr,Weinberg:1967tq,Salam:1968rm}. In such a theory,
the strengths of the Higgs boson couplings to all the elementary particles
(including the Higgs itself) are universally set by the corresponding masses.
A global fit to the production rates, that employs several different final 
states, shows a 10-20\% agreement with SM predictions for the measured
couplings to fermions and to vector bosons~\cite{cms2013,CMS:2014ega,
atlas2013}.  Conversely, no information has yet been obtained on the 
Higgs self-coupling.

To probe elementary couplings in both decay and production is
not only interesting, but also quite useful in order to reduce 
the theoretical as well as the experimental uncertainties.
The case of the Higgs coupling to bottom quarks ($\yb$)
is rather special in this respect: while its strength is significantly
smaller than those relevant to vector bosons and to top quarks, for a 
Higgs mass of about 125 GeV the $H\to b\bb$ mode dominates the total decay 
width, owing to phase-space factors. Unfortunately, this is not particular
helpful from an experimental viewpoint, for several reasons: the total width 
is extremely small in absolute value; the backgrounds which feature
$b$ quarks are immense; and the relative partial decay widths are 
difficult to determine with some accuracy, precisely because the 
$H\to b\bb$ branching ratio is close to one. A Higgs production mode
that features a $\bbH$ coupling is thus a more viable alternative. 
There are two main such modes: the dominant one is a contribution to
gluon fusion, where bottom-quark loop amplitudes interfere
with top-quark loop amplitudes; the second-largest is associated production 
($\bbH$ henceforth) -- with this, one understands processes that at
the Born level receive contributions from tree graphs that include 
a $b$-quark line which radiates a Higgs boson. The $b$-quark contributions
to gluon fusion are of the order of a few percent for the total rate, and up 
to 10\% for a Higgs produced at small transverse momentum (see
e.g.~refs.~\cite{Bagnaschi:2011tu,Mantler:2012bj,Grazzini:2013mca,
Banfi:2013eda,Harlander:2014uea}). Conversely, the fully-inclusive 
$\bbH$ cross section is much smaller: it is of the
same order as the $t\bar{t}H$ one (around 0.5~pb at 13~TeV), i.e.~the fifth largest
after gluon fusion, VBF, $WH$, and $ZH$. However, this inclusive rate 
decreases dramatically when conditions are imposed (which means minimal 
transverse momentum and centrality requirements, typical of $b$-tagging)
that allow one to render it distinguishable from other production mechanisms.

The SM picture outlined above might be significantly modified by 
beyond-SM effects. For example, in extended Higgs sectors the couplings
of the scalar particle(s) to bottom quarks may be enhanced, typically by 
a factor $1/\cos\beta$ or $\tan\beta$ for pseudoscalars in generic or 
supersymmetric 2HDM's. Given that a scalar sector richer than that of 
the SM has not yet been ruled out experimentally, this is a fact that 
one must bear in mind, and that constitutes a strong motivation for 
pursuing the study of scalar-particle production in association
with $b$ quarks.

In addition, and regardless of phenomenological motivations, $\bbH$ 
production is interesting in its own right theoretically, and has in 
fact generated much discussion in the past. The main reason for this
is that, as for all mechanisms that feature $b$ quarks at the hard-process
level, there are two ways of performing the computation, which
are usually called four- and five-flavour schemes (abbreviated
with 4FS and 5FS henceforth, respectively). These two schemes are 
supposed to address issues that arise in different kinematic regimes,
which one may classify by using a hard scale (that we denote by $Q$) 
typical of the process. Lest we complicate the discussion with a 
proliferation of scales, for this discussion we assume to be dealing
with fully-inclusive observables.

Four-flavour scheme computations are relevant to those cases where
the physical mass of the $b$ quark ($\mb\sim 5$~GeV) plays the role
of a hard scale, which means:
\beq
\max\left(\frac{\LQCD}{Q},\frac{\LQCD}{\mb}\right)\ll\frac{\mb}{Q}\,.
\label{4FScond}
\eeq
In the context of a 
factorisation theorem, this implies that the $b$ quark must be
treated as a massive object at the level of the short-distance cross 
section, where it never appears in the initial state; that the factorisation
formula neglects terms of order $\LQCD/\mb$; and that observables with 
tagged final-state $b$ quarks (in association with whatever other object) 
can be computed in perturbation theory. In the case at hand, the leading
order (LO) partonic processes are therefore:
\beq
gg\to b\bb H\,,\;\;\;\;\;\;\;\;
q\bq\to b\bb H\,,
\label{LO4FS}
\eeq
with $q$ a light quark. 

The condition $\mb\sim Q$ is sufficient for eq.~(\ref{4FScond}) to
be satisfied, but it is not necessary. In general, by computing the
cross section in perturbation theory, at any perturbative order the 
4FS result will feature terms proportional to \mbox{$\as^k\log^k(\mb/Q)$}.
These are harmless when $\mb\sim Q$, but if $Q\gg\mb$ they might spoil
the ``convergence'' of the perturbative series, in spite of the condition
in eq.~(\ref{4FScond}) being still fulfilled. This is hardly an unusual
situation, and certainly one which is not peculiar to $b$ physics:
when a series is not well-behaved, one re-organises its expansion,
by resumming the appropriate logarithmic terms. On the other hand,
the relative smallness of $\mb$ renders the presence of large logarithms
a likely occurrence. If the crucial characteristic of an observable is
that of being dominated by such logarithms, then eq.~(\ref{4FScond})
is not really relevant, and a more appropriate description of the
dominant kinematic regime is:
\beq
\max\left(\frac{\LQCD}{Q},\frac{\mb}{Q}\right)\ll\frac{\LQCD}{\mb}\,,
\label{5FScond}
\eeq
which is what the five-flavour scheme computations deal with.
The easiest way to achieve eq.~(\ref{5FScond}), and one that lends
itself particularly well to perturbative computations, is that of
setting $\mb=0$ at the level of short-distance cross sections.
In a factorisation theorem, thus, the $b$ quark will be treated
on equal footing as all of the other light quarks including, in
particular, the fact that it could appear in the initial state.
This implies that, in the 5FS the LO partonic process for Higgs
production in association with $b$ quarks is:
\beq
b\bb\to H\,.
\label{LO5FS}
\eeq
The fact that $\mb=0$ has to be regarded as a technical mean to
achieve the resummation of large logarithms, which 
is the reason why the 5FS has been introduced in the first place.
In the case of a fully-inclusive cross section, the logarithms
are indeed effectively resummed through the Altarelli-Parisi 
evolution equations relevant to the $b$-quark PDF\footnote{Note that
this is not in contradiction with the fact that in the short-distance 
cross section one sets $\mb=0$: logarithms of $\mb/Q$ enter the 
evolution equation through threshold and boundary conditions which,
at variance with the light-flavour case, can be computed in
perturbation theory.}.

From the purely theoretical viewpoint, 5FS computations have the advantage
of being much simpler than their 4FS counterparts: the process in
eq.~(\ref{LO5FS}) is a $2\to 1$ ${\cal O}(\as^0)$ one, while those
in eq.~(\ref{LO4FS}) are $2\to 3$ and of ${\cal O}(\as^2)$. This
advantage comes at a steep price: the calculation associated with
eq.~(\ref{LO5FS}) does provide one with a much more limited information
than eq.~(\ref{LO4FS}), owing to the absence of final-state 
$b$ quarks in the former. In the 5FS, such information can only
be recovered by considering higher orders in $\as$, and in so doing 
the 5FS starts losing the advantage mentioned above. Explicitly for
the case of $\bbH$ production, in the 5FS the leading contribution
is ${\cal O}(\as)$ for $1$-$b$ tag observables, and ${\cal O}(\as^2)$ for 
$2$-$b$ tag observables, while in the 4FS an ${\cal O}(\as^2)$
term is always perturbatively leading, regardless of whether one considers
$0$-, $1$-, or $2$-$b$ tag observables. Furthermore, in the context of a 5FS computation
a $b$-tagged object leads to unphysical results if defined with too small 
a $\pt$ (unphysical since the $b$-tagged cross section is larger than the 
fully-inclusive one; it diverges at $\pt\to 0$), owing to the mass of the 
$b$ quark having been set equal to zero; this problem does not affect 4FS 
results. Finally, $b$-tagged objects in the 5FS cannot coincide with
$b$ quarks (which, conversely, is the case in the 4FS), because the
corresponding cross section is not finite order-by-order in perturbation 
theory: the $b$ quarks must either be integrated over, or be part of jets, 
or be converted into $b$-flavoured hadrons through the convolution 
with fragmentation functions.

One must also bear in mind that, in general, when considering differential 
observables new mass scales become relevant to the problem, which rapidly
render the generalisation of eqs.~(\ref{4FScond}) and~(\ref{5FScond})
impractical, and make it difficult to decide a priori which calculational
scheme is best suited to tackle the problem at hand.

While 4FS results lack logarithmic terms beyond the first few, 5FS results
lack power-suppressed terms $(\mb/Q)^n$. Which of the two classes of
terms is more important depends on the observable studied, that determines
the dominant kinematic regime. In order to avoid the problems that
this fact entails, schemes~\cite{Aivazis:1993pi,Thorne:1997ga,Thorne:2006qt,
Cacciari:1998it,Kramer:2000hn,Tung:2001mv} have been proposed 
that allow one to combine, possibly in a systematic manner in 
perturbation theory, the logarithmic and 
power-suppressed terms in a single cross section, which is appropriate
to all kinematic regimes.\footnote{A pragmatic approach to combine the total 
inclusive cross sections in the 4FS and 5FS was proposed in ref.~\cite{Harlander:2011aa}.}
 It is customary (but not necessary) to view these 
approaches as power-terms-improved 5FS calculations, which is sort of 
natural when one looks at fairly inclusive observables, for which one 
expects logarithms to be more important than power-suppressed
terms, so that the latter are small corrections to the former.
In the following and for the sake of clarity, by 5FS results we shall 
understand those that, at the level of short-distance cross sections, 
do not contain any power-suppressed terms.

Given what has been said so far, a typical rationale is the following: 
if logarithms are large, the 5FS should be superior to the 4FS; if
they are not, and thus power-suppressed terms might be important, 
then 4FS approaches should be preferred. Unfortunately, the relative
weight of logarithmic and power-suppressed terms is observable dependent
(which is also complicated by the fact that a given observable
can be potentially associated with different powers of $\as$ in the 
four- and five-flavour schemes, as discussed above). However, one
expects that, for processes and in regions of the phase space where 
both resummation and mass effects are not dominant, the two approaches 
should give similar results. As a matter of fact, at least for
inclusive quantities 5FS and 4FS results are indeed generally similar (in 
particular, for $\bbH$ production), because of the two following
facts. {\em i)} Logarithms are dominant, but they are especially so
only at large Bjorken $x$'s, and are always associated with
phase-space suppression factors~\cite{Maltoni:2012pa}. {\em ii)} 
At the lowest perturbative orders (LO and NLO), a reasonable agreement
is found only by judiciously choosing the hard scales; in particular,
arguments based on collinear dominance suggest that the optimal values 
of these scales in $\bbH$ production are significantly smaller 
than $\mH$~\cite{Maltoni:2003pn,Boos:2003yi,Harlander:2003ai}, 
i.e.~than the hardness one would naively associate with this 
production process. As a consequence of these two facts, the few
logarithms that appear in a 4FS fixed-order cross section approximate well 
numerically the leading logarithmic tower(s) present in the corresponding 5FS
cross section (especially at the NLO), while power-suppressed terms are 
unimportant. There is an ample heuristic evidence of the facts above. 
However, one must keep in mind that such an evidence, and the arguments that
support it, are chiefly relevant to inclusive variables. In a more
exclusive scenario, it is important to assess the situation in an
unbiased manner: this is one of the main aims of this paper.

What has been said so far has tacitly assumed fixed-order parton-level
computations; when matrix elements are matched to parton showers (PS),
some aspects of the picture do change. Let us start by considering
5FS approaches. For these, the foremost consequence of PS matching is the 
fact that even the ${\cal O}(\as^0)$ cross section of eq.~(\ref{LO5FS}), 
thanks to the backward evolution of the initial-state $b$'s, will generate 
$b$-flavoured hadrons in the final state, which will render realistic any 
$b$-tagging requirements, regardless of the $\pt$ values where they are 
imposed. While this goes a long way towards improving 5FS 
predictions, one must not forget two facts. Firstly, it is well known 
that the backward evolution of massless $b$ quarks is not trivial 
for Monte Carlos (MCs; see e.g.~sect.~3.3 of ref.~\cite{Frixione:2010ra},
and later here in sect.~\ref{sec:4FSvs5FS});
this can lead to unexpected results for certain classes of observables,
and to a marked MC dependence. In particular, the necessary kinematic 
reshuffling of the massless $b$ quarks into massive $b$ quarks can have 
a significant impact on the kinematics of final-state $B$ mesons. 
Secondly, beyond LL the Altarelli-Parisi
equations and the MC backward evolution do differ (and especially so
for the treatment of $b$-quark thresholds); although generally small, 
these differences might become relevant in certain phase-space corners, 
where comparisons to data will help decide which description is best.
As far as 4FS predictions are concerned, they are also improved by their 
matching with PS. In particular, small-$\pt$ initial-state emissions are
Sudakov-suppressed, and effectively resummed by the parton 
showers\footnote{This is not the same as the resummation performed in 
the 5FS case; in a
PS-matched 4FS, it remains true that no initial-state $b$ quark is
present which could be backward evolved.}. Note, finally, that the 
matching to PS introduces in both the 4FS and 5FS extra power-suppressed
effects, due to long-distance phenomena.

In summary, there are a number of interesting physics questions
that are relevant to $\bbH$ production, which become especially
crucial when exclusive quantities are studied, in particular
in the context of computations matched to parton showers.
The main results of this paper are the following.
\begin{itemize}
\item We present the first NLO computations matched to parton
showers (NLO+PS) in the four- and five-flavour schemes. At the level of 
fixed-order NLO results (fNLO) we present fully-differential results 
which extend the scope of those available in the 
literature~\cite{Dittmaier:2003ej,Dawson:2003kb} in a very 
significant manner.
\item We study, for the first time, the effect of the ${\cal O}(\yb\yt\as^3)$ 
interference term on differential distributions in the 4FS,
in particular at the NLO+PS accuracy.
\item We compare 4FS and 5FS NLO+PS predictions at the level of differential
distributions, in order to further the arguments given above for such
observables. For the inclusive Higgs transverse momentum, we
also compare to the (N)NLO+(N)NLL analytical results of
ref.~\cite{Harlander:2014hya}.
\end{itemize}
All of our computations, bar those that feature analytical resummations, 
are performed in the automated \aNLO\ framework~\cite{Alwall:2014hca}.

The paper is organised as follows: in sect.~\ref{sec:outline} we 
report some generalities relevant to our 4FS and 5FS computations; 
phenomenological results are presented in sect.~\ref{sec:results} -- see 
in particular sect.~\ref{sec:4FS} for 4FS predictions, and 
sect.~\ref{sec:4FSvs5FS} for 4FS-vs-5FS comparisons; 
we conclude in sect.~\ref{sec:conclusions}.

\section{Outline of the calculations\label{sec:outline}}
In this section, we briefly describe the physics contents 
relevant to our 4FS and 5FS predictions which have been calculated 
with \aNLO.
We remind the reader that \aNLO\ contains all ingredients relevant
to the computations of LO and NLO cross sections, with or without
matching to parton showers. NLO results not matched to parton showers
(called fNLO~\cite{Alwall:2014hca}) are obtained by adopting the
FKS method~\cite{Frixione:1995ms,Frixione:1997np} for the subtraction 
of the singularities of the real-emission matrix elements (automated 
in the module \MadFKS~\cite{Frederix:2009yq}), and the OPP
integral-reduction procedure~\cite{Ossola:2006us} for the computation 
of the one-loop matrix elements (automated in the module 
\MadLoop~\cite{Hirschi:2011pa}, which makes use of
\CutTools~\cite{Ossola:2007ax} and of an in-house implementation 
of the optimisations proposed in ref.~\cite{Cascioli:2011va} (\OL)).
Matching with parton showers is achieved
by means of the MC@NLO formalism~\cite{Frixione:2002ik}.
\aNLO\ is maximally automated, since the only operations required
by the user are to enter the process he/she is interested in generating,
and the associated input parameters. The case of $\bbH$ production 
has one peculiarity related to the central role of the bottom Yukawa, 
which could not be handled by the public version available when 
ref.~\cite{Alwall:2014hca} was released, and which has been the object of a
special treatment for the sake of this paper; more details are given in the
next section.

Before introducing the features of our own computations, we briefly
summarise the status of the $\bbH$-production results available in the
literature. As far as the 4FS is concerned, NLO fixed-order parton-level
predictions have been presented in refs.~\cite{Dittmaier:2003ej,
Dawson:2003kb}, and later updated to the case of MSSM-type 
couplings~\cite{Dawson:2005vi}, and to SUSY-QCD corrections
in the MSSM~\cite{Liu:2012qu,Dittmaier:2014sva}. The focus of these papers
is the total cross section; only a handful of differential predictions
have been shown. The literature is considerably richer for the 5FS,
owing to its being technically simpler. NLO and NNLO QCD corrections 
for total rates were first computed in refs.~\cite{Dicus:1998hs,Balazs:1998sb}
and in ref.~\cite{Harlander:2003ai}, respectively.\footnote{Even the ingredients 
for the full N$^3$LO prediction are already available \cite{Ahmed:2014pka,Gehrmann:2014vha};
 their combination is far from trivial though.}
Differential parton-level predictions have been later made available:
at the NLO for $H+b$ and $H+$jet production~\cite{Campbell:2002zm,
Harlander:2010cz}, and at the NNLO for jet rates~\cite{Harlander:2011fx}
and fully differential distributions~\cite{Buehler:2012cu}. The Higgs 
transverse momentum has been studied analytically at the NNLO in 
ref.~\cite{Ozeren:2010qp}, while resummed NLO+NLL and NNLO+NNLL results 
have been presented in ref.~\cite{Belyaev:2005bs} and 
ref.~\cite{Harlander:2014hya}, respectively.
Prior to this paper, no NLO+PS predictions have been obtained
in either scheme. 

\subsection{Four-flavour scheme\label{sec:out4FS}}
\begin{figure}[!ht]
\begin{center}
  \subfloat[\label{fig:4FSLO:a}]{
    \includegraphics[width=0.3\textwidth]{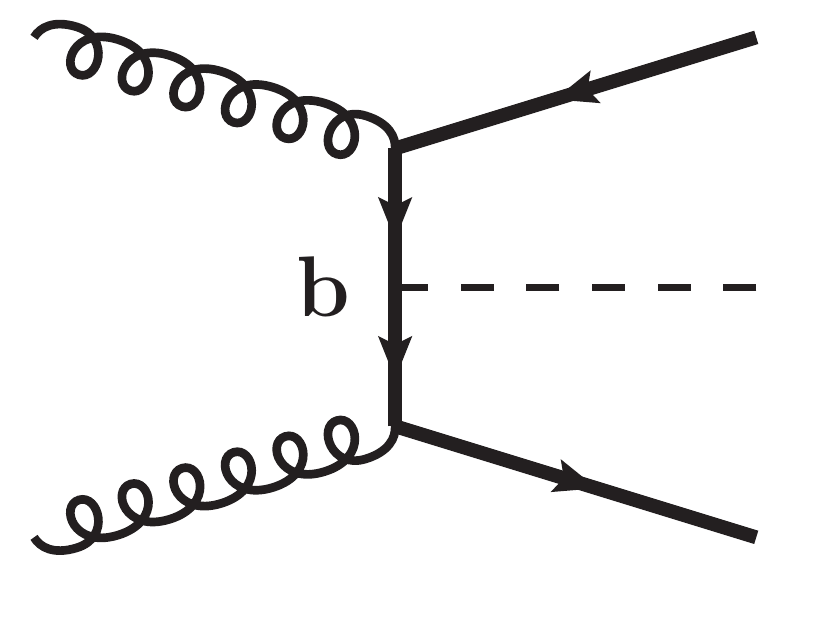}
  }
$\phantom{aaaaaaa}$
  \subfloat[\label{fig:4FSLO:b}]{
    \includegraphics[width=0.3\textwidth]{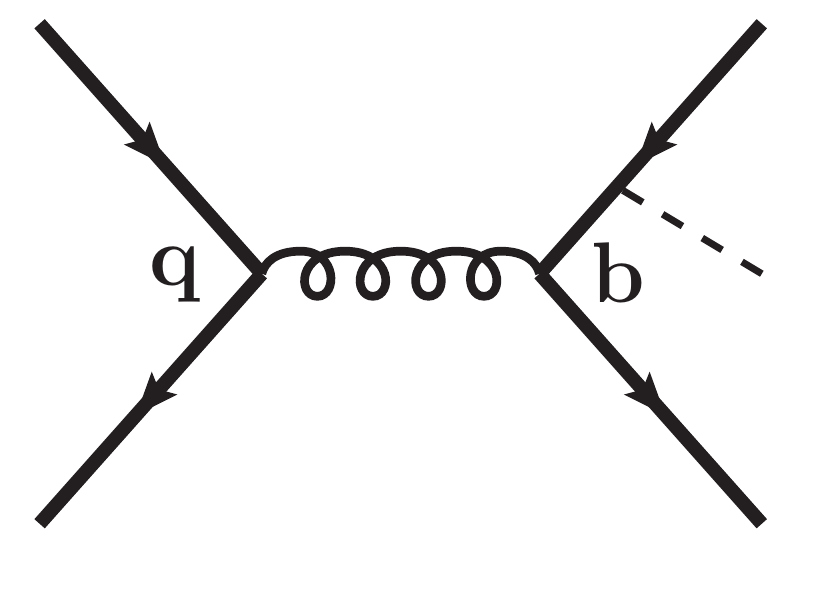}
  }
  \caption{Sample of LO Feynman diagrams for $\bbH$ production in
the four-flavour scheme, for the two relevant classes of partonic 
subprocesses: (a) $gg\rightarrow b\bar{b}H$; 
(b) $q\bar{q}\rightarrow b\bar{b}H$.}
  \label{fig:4FSLO}
\end{center}
\end{figure}

At the LO, the 4FS cross section receives contributions from
the ${\cal O}(\as^2)$ $2\to 3$ partonic subprocesses given in
eq.~(\ref{LO4FS}); representative Feynman diagrams are displayed
in fig.~\ref{fig:4FSLO}. The Higgs is always radiated off a 
$b$ quark, and therefore the cross section at this perturbative
order is proportional to $\yb^2\as^2$. 

\begin{figure}[!ht]
\begin{center}
\hspace{-0.46cm}
  \subfloat[\label{fig:4FSoneloop:a}]{
    \includegraphics[width=0.28\textwidth]{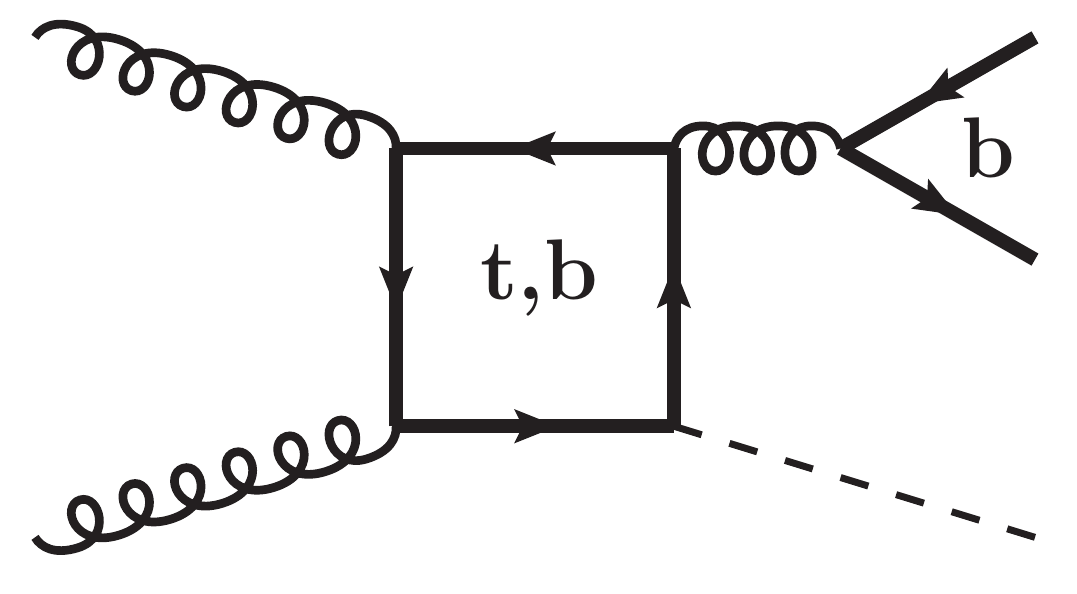}
  }
$\phantom{..}$
  \subfloat[\label{fig:4FSoneloop:b}]{
    \includegraphics[width=0.28\textwidth]{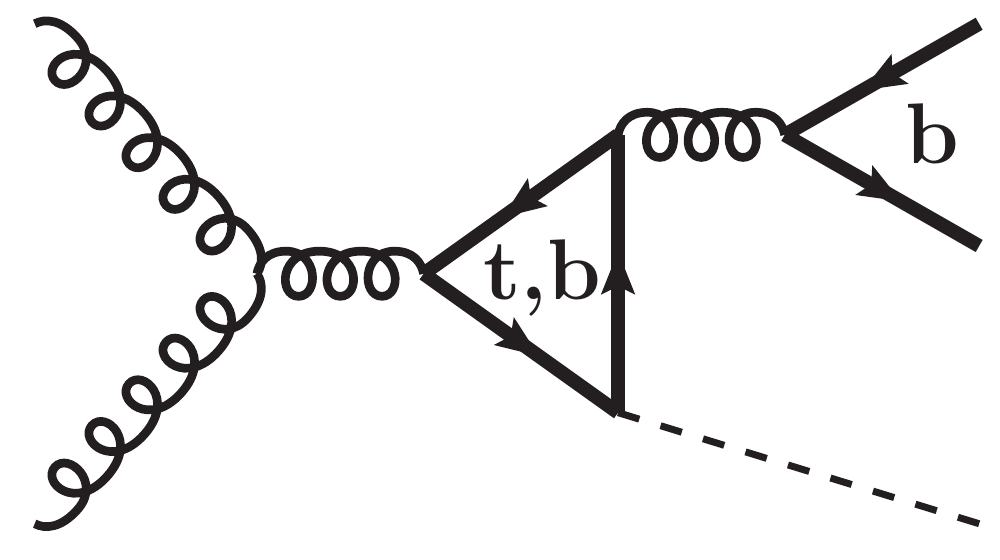}
  }
$\phantom{aaa}$
  \subfloat[\label{fig:4FSoneloop:c}]{
    \includegraphics[width=0.24\textwidth]{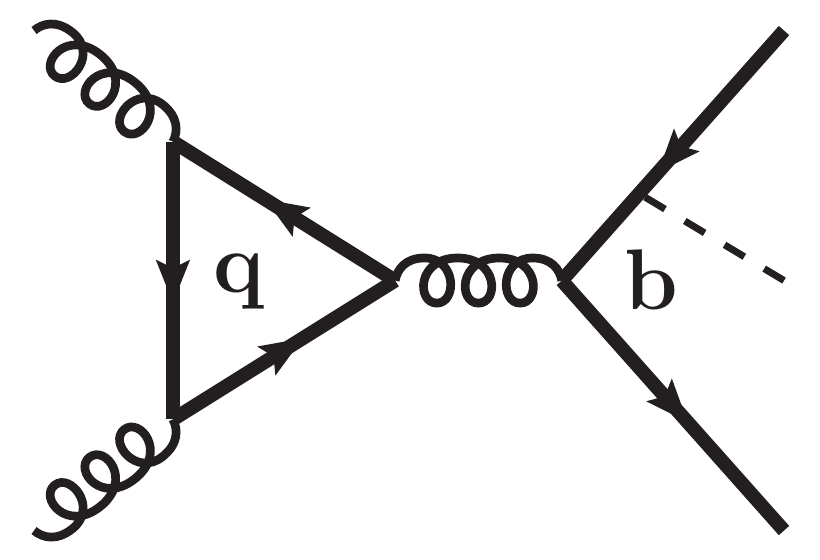}
  }
\\
  \subfloat[\label{fig:4FSoneloop:d}]{
    \includegraphics[width=0.28\textwidth]{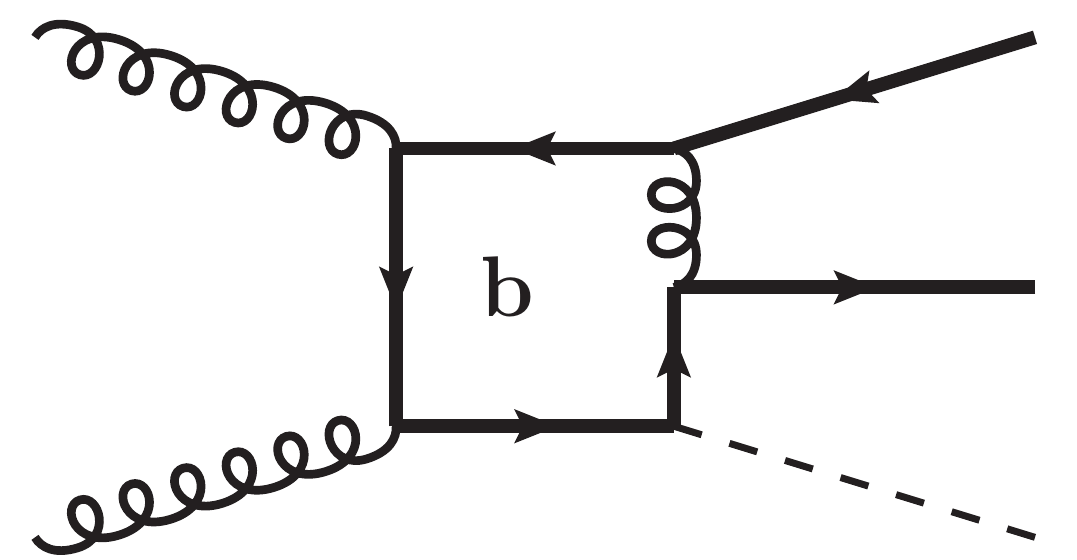}
  }
$\phantom{aa}$
  \subfloat[\label{fig:4FSoneloop:e}]{
    \includegraphics[width=0.26\textwidth]{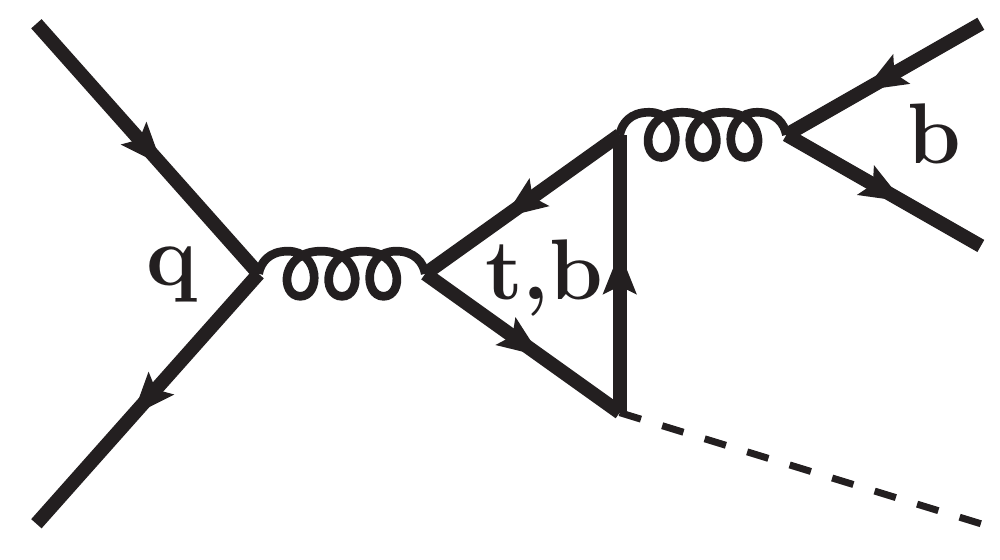}
  }
$\phantom{aa}$
  \subfloat[\label{fig:4FSoneloop:f}]{
    \includegraphics[width=0.28\textwidth]{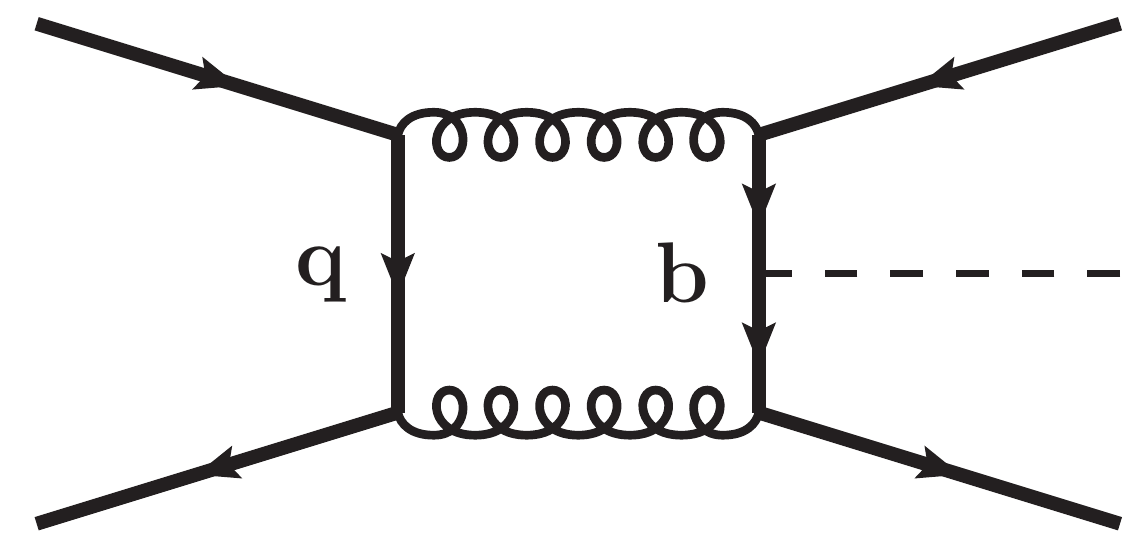}
  }
  \caption{Sample of one-loop Feynman diagrams for $\bbH$ production in the
four-flavour scheme. All diagrams contribute to the $\yb^2$ term when the
Higgs is radiated off a bottom quark, while diagrams (a), (b), and (e) with a
top-quark loop contribute to the $\yb\yt$ term.}
  \label{fig:4FSoneloop}
\end{center}
\end{figure}

\begin{figure}[!ht]
\begin{center}
\hspace{-0.5cm}
  \subfloat[\label{fig:4FSreal:a}]{
    \includegraphics[width=0.3\textwidth]{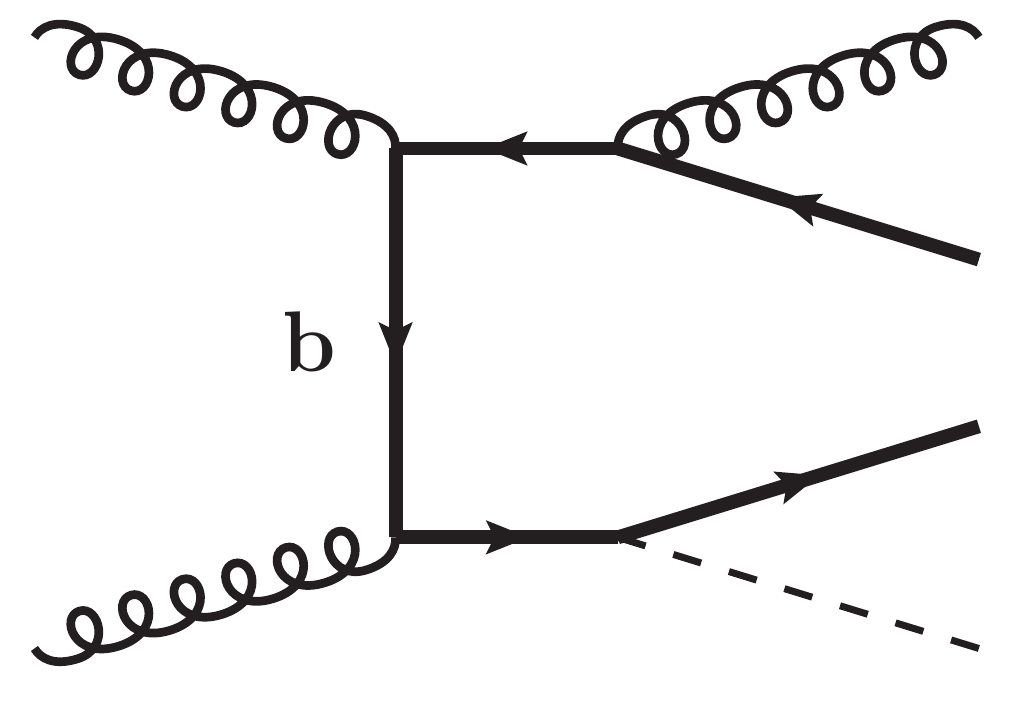}
  }
$\phantom{aaaaaaa}$
  \subfloat[\label{fig:4FSreal:b}]{
    \includegraphics[width=0.3\textwidth]{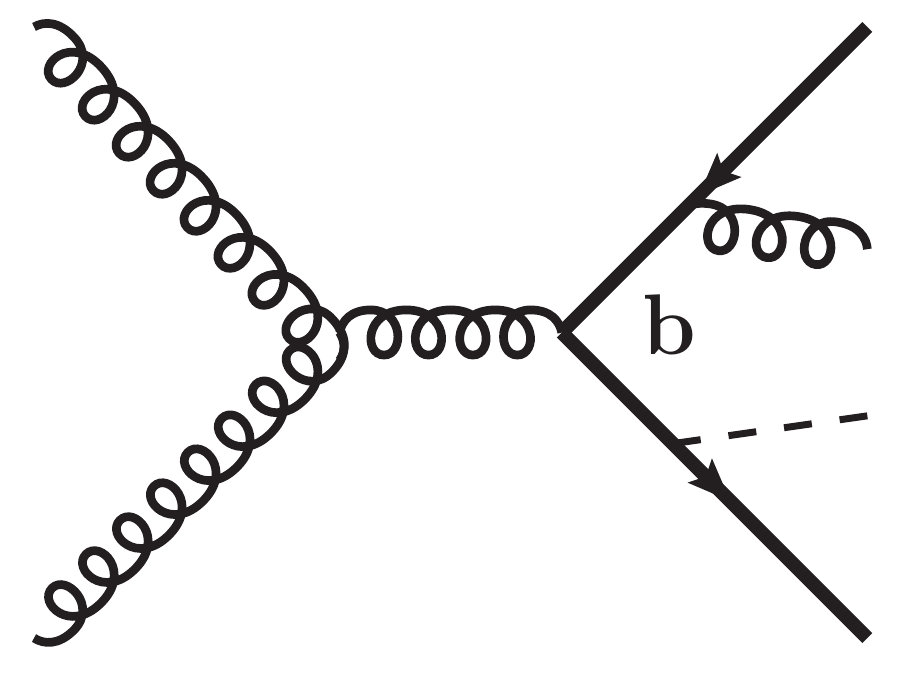}
  }
\\
  \subfloat[\label{fig:4FSreal:c}]{
    \includegraphics[width=0.25\textwidth]{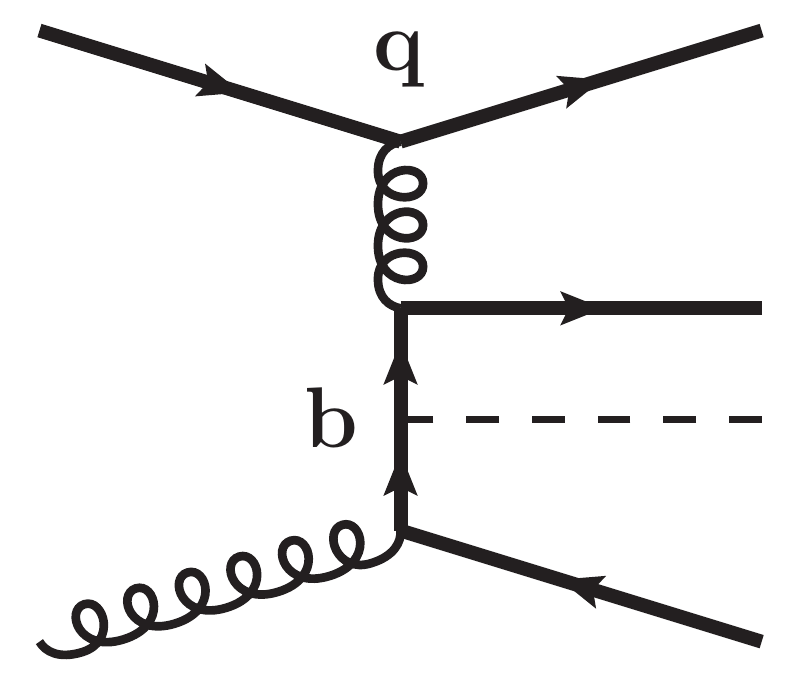}
  }
$\phantom{aaaaaaaaa}$
  \subfloat[\label{fig:4FSreal:d}]{
    \includegraphics[width=0.3\textwidth]{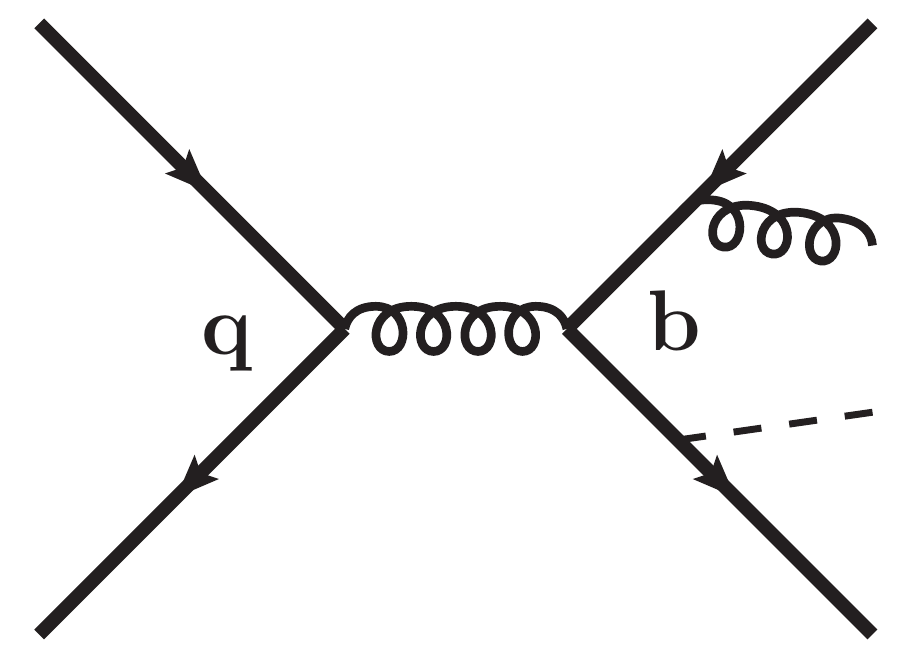}
  }
  \caption{Sample of real-emission Feynman diagrams for $\bbH$ production in
the four-flavour scheme.}
  \label{fig:4FSreal}
\end{center}
\end{figure}

The coupling structure becomes more involved when one considers
NLO corrections. As is well known, these can be classified as being of either
virtual or real-emission origin; sample diagrams for these two classes
are displayed in fig.~\ref{fig:4FSoneloop} and fig.~\ref{fig:4FSreal}
respectively. Consider, in particular, the virtual diagrams of
fig.~\ref{fig:4FSoneloop:a}, \ref{fig:4FSoneloop:b}, 
and~\ref{fig:4FSoneloop:e}: when the heavy quark that circulates
in the loop is a top, the corresponding amplitude is proportional
to $\yt$, and does not feature the bottom Yukawa $\yb$. All of 
the other diagrams, as well as those relevant to real emissions,
have amplitudes proportional to $\yb$. At the NLO, the cross section
receives contributions from the interference of the one-loop diagrams
with the Born ones, and from the squares of the real-emission diagrams.
The squares of the one-loop diagrams, in turn, will enter the NNLO
result. One can thus write the $\bbH$ 4FS cross section up to
${\cal O}(\as^4)$ as follows:
\beq
\sigma_{\bbh}^{\fs{4}}=
\underbrace{\as^2\,\yb^2\,\Delta^{(0)}_{\yb^2}+
\as^3\,\Big(\yb^2\,\Delta^{(1)}_{\yb^2}}_{\equiv \sigma_{\yb^2}}+
\underbrace{\yb\,\yt\,\Delta^{(1)}_{\yb\yt}}_{\equiv \sigma_{\yb\yt}}\Big)+
\as^4\left(\yb^2\,\Delta^{(2)}_{\yb^2} + \yb\,\yt\,\Delta^{(2)}_{\yb\yt} +
\yt^2\,\Delta^{(2)}_{\yt^2}\right)\,.
\label{sig4FS}
\eeq
The $\bbH$ NLO results presented in the literature focus on
the $\sigma_{\yb^2}$ term of eq.~(\ref{sig4FS}). We are not aware 
of existing predictions for $\sigma_{\yb\yt}$ at the level of differential
observables, whose impact we shall discuss in 
sect.~\ref{sec:4FS}. Finally, all terms of ${\cal O}(\as^4)$
have been ignored here; note that at least those proportional to
$\yt^2$ are usually seen as NNLO contributions to the gluon-fusion
cross section.

The fully automated \aNLO~\cite{Alwall:2014hca} program can handle
4FS $\bbH$ production rather straightforwardly -- the calculation is of
a complexity similar to that of $Zb\bb$ production, which could be
studied~\cite{Frederix:2011qg} even with a version of the code much
less powerful than the present one. However, the default treatment
of Yukawa couplings in the code is that of an on-shell scheme 
renormalisation, which is not optimal in the case of $\bbH$ production, 
where the $\msbar{}$ scheme has to be preferred~\cite{Braaten:1980yq}.
Such a scheme is indeed what has been employed in previous NLO 4FS 
computations~\cite{Dittmaier:2003ej,Dawson:2003kb}, since the use
of an $\msbar{}$ renormalized Yukawa $\overline{y}_b(\muR)$ has the
advantage of resumming to all orders potentially large logarithms
of \mbox{$\mH/\mb$}, when \mbox{$\muR\sim\mH$} is chosen.
A change in the renormalisation scheme, and the UV counterterms 
it entails, is simply dealt with at the level of the \UFO\ 
model~\cite{Degrande:2011ua} that \aNLO\ has to import prior
to generating a process (see appendix~B.1 of ref.~\cite{Alwall:2014hca}
for further details), and by including the relevant routines that perform
the running of $\overline{m}_b(\mu)$. There is only one extra complication, due to
the fact that the $\msbar{}$ Yukawa introduces in the cross section
an extra $\muR$ dependence w.r.t.~those taken explicitly into account
in ref.~\cite{Frederix:2011ss}, which are used by the code for the
definition of scale- and PDF-independent coefficients that are exploited
for the a-posteriori computation of scale and PDF uncertainties
by means of reweighting. Furthermore, such a dependence is different
in the $\sigma_{\yb^2}$ and $\sigma_{\yb\yt}$ terms introduced in
eq.~(\ref{sig4FS}), owing to the different powers of $\yb$ that
appear in those terms. Although this complication will become
recurrent in a mixed-coupling expansion scenario, at the moment it
does not warrant a completely general and automated solution.  
Therefore, we have treated $\bbH$ production as a special case,
by integrating separately the $\sigma_{\yb^2}$ and $\sigma_{\yb\yt}$ 
terms, which necessitate loop-content filtering operations (see sect.~2.4.2
of ref.~\cite{Alwall:2014hca}) in order to exclude, or to include
only, top-quark loops in the virtuals. For each of these two terms,
we have manually performed the modifications in
the definition of the coefficients, mentioned above, that serve
to compute the theoretical systematics. Apart from these manipulations,
the generation and subsequent computation of the $\bbH$ 4FS cross section
proceed exactly with the same general steps as those described
in ref.~\cite{Alwall:2014hca}, namely:

\noindent
~~\prompt\ {\tt ~import model loop\_sm\_MSbar\_yb}

\noindent
~~\prompt\ {\tt ~generate p p > h b b\~{} [QCD]} 

\noindent
followed by the standard {\tt output} and {\tt launch} commands,
and where {\tt loop\_sm\_MSbar\_yb} is the name of the \UFO\ model 
that includes the appropriate UV counterterms for the renormalisation 
of the bottom Yukawa in the $\msbar{}$ scheme.

Given that the few manual operations mentioned above are necessary on top 
of the commands just listed, the user interested in the simulation of
$\bbH$ production with \aNLO\ is strongly encouraged to contact us.

\begin{figure}[t]
\begin{center}
  \subfloat[\label{fig:5FS:a}]{
    \includegraphics[width=0.28\textwidth]{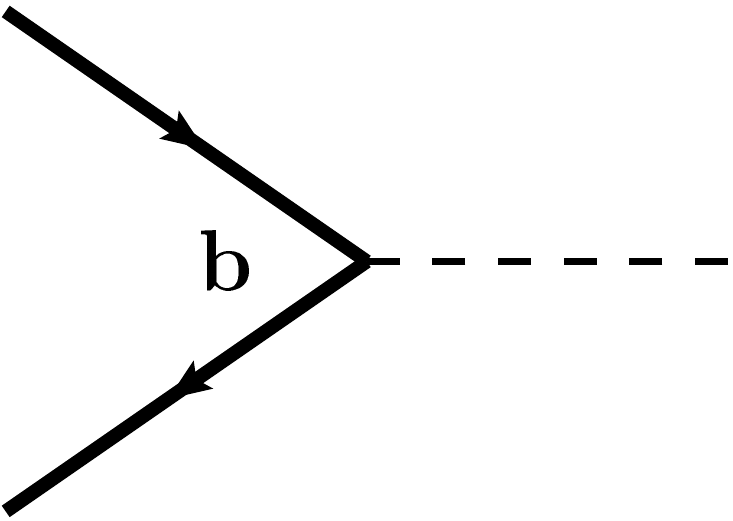}
  }
$\phantom{aaaaaaaa}$
  \subfloat[\label{fig:5FS:b}]{
    \includegraphics[width=0.28\textwidth]{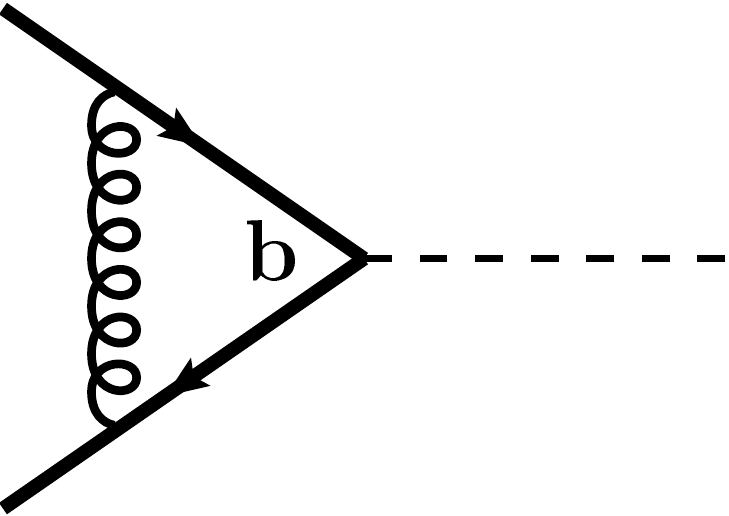}
  }
\\
  \subfloat[\label{fig:5FS:c}]{
    \includegraphics[width=0.28\textwidth]{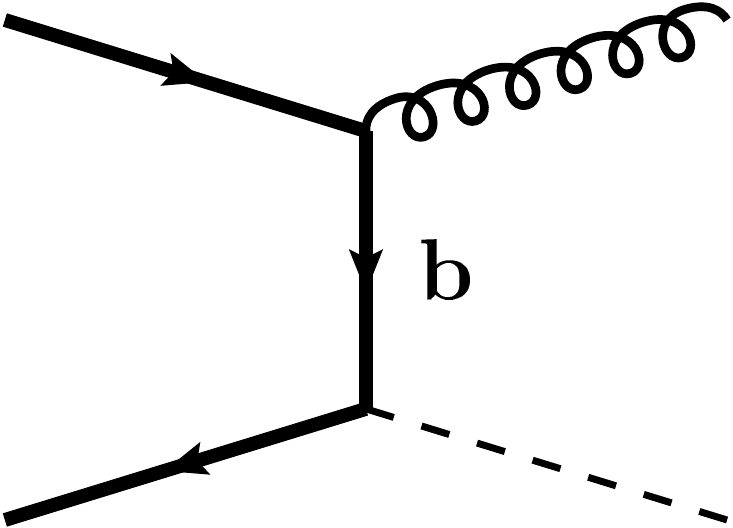}
  }
$\phantom{aaaaaaaa}$
  \subfloat[\label{fig:5FS:d}]{
    \includegraphics[width=0.28\textwidth]{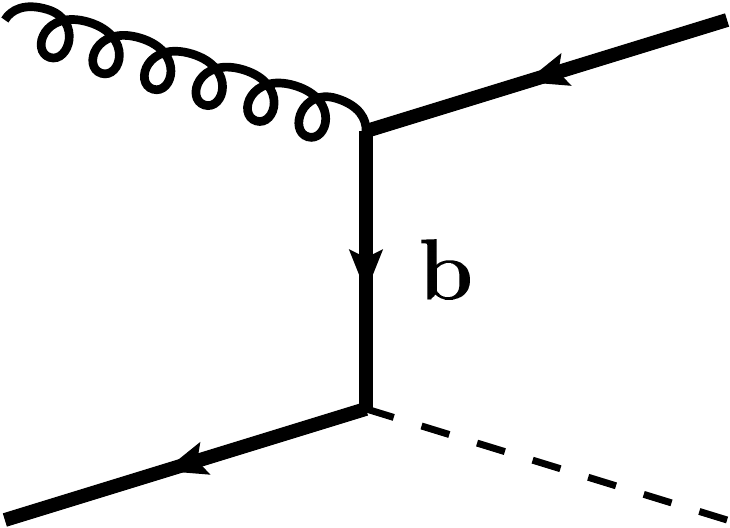}
  }
  \caption{A sample of Feynman diagrams for $\bbh{}$ production in
   the five-flavour scheme: (a) \lo{}; (b) one-loop; (c-d) real emission.}
  \label{fig:5FS}
\end{center}
\end{figure}
\subsection{Five-flavour scheme\label{sec:out5FS}}
Consistently with the general discussion given in sect.~\ref{sec:intro},
our 5FS results are obtained by setting $\mb=0$ (while keeping the 
bottom Yukawa finite). Sample Feynman diagrams that contribute
to the cross section in this scheme are displayed in fig.~\ref{fig:5FS:a}
(at the LO, ${\cal O}(\as^0)$) and fig.~\ref{fig:5FS:b}--\ref{fig:5FS:d} 
(at the NLO, ${\cal O}(\as)$); the cross section is proportional
to $\yb^2$, and the $\yb\yt$ term is absent at the $\as$ order at 
which we are working. Analogously to what has been done in the 4FS, 
the bottom Yukawa is renormalised in the $\msbar{}$ scheme 
(see sect.~\ref{sec:out4FS}).
The general comments made before that concern the generation
of the process apply to the 5FS case as well. At variance with
the 4FS calculation, however, the imported model must have the
$b$ quark mass set equal to zero (except in the Yukawa coupling). 
The relevant \aNLO\ commands are thus:

\noindent
~~\prompt\ {\tt ~import model loop\_sm\_MSbar\_yb-no\_b\_mass}

\noindent
~~\prompt\ {\tt ~define p = g u d s c b u\~{} d\~{} s\~{} c\~{} b\~{}}

\noindent
~~\prompt\ {\tt ~generate p p > h [QCD]} 

\noindent
where the second line explicitly instructs the code to consider
the $b$ quark as part of the proton. Note that since the imported model
is the Standard Model, and not the effective theory which features
an effective $ggH$ vertex, the {\tt generate} command will indeed
result in creating the 5FS $\bbH$ cross section we are interested into.

\section{Phenomenological results\label{sec:results}}
In this section we present several differential distributions
that we reconstruct from final-state particles in $\bbH$ production
at the 13~TeV LHC. Although we work in the SM, our predictions 
are directly applicable to $b\bar{b}\phi$ production (with a neutral 
$\phi=h,H,A$) in a \thdm{}-type extension of the SM, by an appropriate 
rescaling of the bottom Yukawa; in the case of the MSSM, this has been 
verified~\cite{Dittmaier:2006cz,Dawson:2011pe}
to be an excellent approximation of the complete result.

Our reference predictions are (N)LO+PS-accurate; where appropriate, we 
also show f(N)LO and (N)NLO+(N)NLL results. For simulations matched
to parton showers, we employ both \HWpp~\cite{Bahr:2008pv,Bellm:2013lba} 
and \PYe~\cite{Sjostrand:2007gs}; for further information on the calculation 
of the MC counterterms relevant to the MC@NLO method, see 
refs.~\cite{Frixione:2010ra,Torrielli:2010aw}.
Higgs decays have not been considered; in particular, the contents
of all jets in the events are solely due either to hard-process particles,
or to radiation off those particles.

\subsection{Input parameters\label{sec:input}}
The central values of the renormalisation ($\muR$) and factorisation 
($\muF$) scales for all \aNLO\ runs ((N)LO+PS and f(N)LO) have been taken 
equal to the reference scale:
\beq
\mu=\frac{\Ht}{4}\equiv 
\frac{1}{4}\sum_i\sqrt{m_i^2+\pt^2(i)}\,,
\label{scref}
\eeq
where the sum runs over all final-state particles at the hard-process
level; this is in keeping with the findings of refs.~\cite{Maltoni:2003pn,
Boos:2003yi,Harlander:2003ai,Maltoni:2012pa}. The theoretical uncertainties 
due to the $\muR$ and $\muF$ dependencies have been evaluated by varying 
these scales independently in the range:
\beq
\frac{1}{2}\mu\le\muR,\muF\le 2\mu\,.
\label{scalevar}
\eeq
The calculation of this theory systematics does not entail any 
independent runs, and is performed by means of the reweighting technique 
introduced in ref.~\cite{Frederix:2011ss}, with the $\bbH$-specific
upgrade discussed in sect.~\ref{sec:out4FS}. In the case of the
analytically-resummed cross sections, we have set the reference
scale equal to:
\beq
\mu_{\sss AR}=\frac{\mt(H)}{4}\equiv 
\frac{1}{4}\sqrt{\mH^2+\pt^2(H)}\,.
\label{screfAR}
\eeq
The so-called resummation scale ($\Qres$) in analytic transverse momentum
resummation plays the role of a matching scale between the low- and high-$\pt$
regions. In order to optimise the high-$\pt$ matching of the resummed to the
fixed-order cross section, this scale should be set equal to about half of
the factorisation scale~\cite{Harlander:2014hya}, and therefore we choose
$\Qres=\mH/8$ as our default value; note that at the NNLO+NNLL, the choice 
of the resummation scale has hardly any impact in the small-$\pt$ region.

We have adopted the MSTW2008 PDF set~\cite{Martin:2009iq}, with its
associated $\as$ value, in its four- or five-flavour variant in agreement
with the 4FS or 5FS computation being carried out. In the case of LO+PS 
simulations, LO PDFs and one-loop $\as$ have been used.
The mass of the Higgs is $\mH=125$~GeV.
The $\msbar{}$ bottom Yukawa, which we compute at the scale $\muR$, is derived 
beyond the LO from the input value $\overline{m}_b(\overline{m}_b)=4.34$~GeV 
in the 4FS, and from $\overline{m}_b(\overline{m}_b)=4.16$~GeV in the 5FS;
these values correspond to a pole mass $\mbottom^{\text{\os{}}}=4.75$~GeV 
at the NLO and NNLO, respectively, and are chosen\footnote{Note that the
value of $\overline{m}_b$ in the 5FS merely affects the normalisation, rather 
than the shapes which are the primary subject of our comparison to the 4FS.}
 for consistency with
the most recent and accurate results for total cross sections in
either scheme~\cite{Dittmaier:2011ti,LHCHXSWG}.
At the LO, since
$\overline{m}_b(\overline{m}_b)=\mbottom^{\text{\os{}}}$, we use
$\overline{m}_b(\overline{m}_b)=4.75$~GeV. In the 4FS, the internal 
bottom and top masses are taken equal to $\mb^{\text{\os{}}}=4.75$~GeV 
and $\mtop^{\text{\os{}}}=173$~GeV, respectively.
Jets are reconstructed with the anti-$\kt$ algorithm~\cite{Cacciari:2008gp},
as implemented in {\sc\small FastJet}~\cite{Cacciari:2011ma}, with a jet radius
of $R=0.5$, and subject to the condition $\pt(j)\ge 25$~GeV. In the case of 
(N)LO+PS simulations, jets are made up of hadrons; $b$-jets (defined to
be jets that contain at least one $B$ hadron (at (N)LO+PS) or $b$ quark 
(at f(N)LO)) are kept only if they fulfill the extra condition 
$\abs{\eta(j_b)}\le 2.5$. We have not generated underlying events.

\subsection{Four-flavour scheme results\label{sec:4FS}}
In this section we present 4FS results for total rates, possibly within 
cuts, and for differential distributions constructed with the 
four-momenta of the Higgs, $b$-jets, and $B$ hadrons or $b$ quarks.
Before giving definite predictions for those quantities, however,
there is a general issue that we would like to address, directly
related to the novelty of being able to perform 4FS computations
at the NLO+PS accuracy. This issue stems from a general characteristics
of $\bbH$ production mentioned in sect.~\ref{sec:intro}, namely the
fact that the optimal values for the hard scales that enter the calculation
appear to be significantly smaller than the hardness of the process
would suggest, which is what has led us to the setting of 
eq.~(\ref{scref}). When one considers parton showers,
another hard scale becomes relevant, which loosely speaking can be
identified with the largest hardness accessible to the shower;
let us denote this scale by $\Qshow$. It is the MC that 
determines, event-by-event, the value of $\Qshow$, by choosing it
so as to maximise the kinematic population of the phase-space due to
shower radiation, without overstretching the approximations upon which 
the MC is based. The latter condition typically implies that
\beq
\langle\Qshow\rangle\,\lesssim\,\mu\,,
\label{Qshrange}
\eeq
where the average is taken over all generated events.
In the context of the MC@NLO method, the condition of eq.~(\ref{Qshrange})
is actually not so crucial: essentially, MC hard radiation is subtracted,
and replaced by that of matrix element origin, and in this way its impact
on physical observables is suppressed. It is however not identically
equal to zero, because the subtraction works within approximation, i.e.~at
the NLO; hence, MC hard radiation formally of NNLO and beyond can still
contribute to the cross section. In order to assess this higher-order
systematics of MC origin (which in MC@NLO is tantamount to the
matching systematics), in \aNLO\ one is given the possibility of
setting the value\footnote{If the MC-determined $\Qshow$ value 
is lower than that set by the user, the latter is ignored. Also
bear in mind that the physical meaning of $\Qshow$ depends on the
specific MC employed -- see ref.~\cite{Alwall:2014hca}.}
of $\Qshow$; this value is actually picked up at random in a 
user-defined range:
\beq
\alpha f_1\sqrt{s_0}\le\Qshowmax\le\alpha f_2\sqrt{s_0}\,,
\label{murange}
\eeq
so as to avoid possible numerical inaccuracies due to the presence 
of sharp thresholds; more details can be found in sect.~2.4.4 of
ref.~\cite{Alwall:2014hca} (see in particular eq.~(2.113) and the
related discussion). In eq.~(\ref{murange}), $s_0$ is the Born-level
partonic c.m.~energy squared, and $\alpha$, $f_1$, and $f_2$ are numerical
constants whose defaults are $1$, $0.1$, and $1$ respectively. 
The way in which $\Qshowmax$ is generated results in a 
distribution peaked at values slightly larger than 
\mbox{$\alpha(f_1+f_2)\sqrt{\langle s_0\rangle}/2$}.
The essence of the MC@NLO method is such that, in practice,
virtually all processes studied so far exhibit a modest
systematics due to the parameters that control $\Qshowmax$.

It is clear by construction that there is a rather direct relationship 
between $\Qshowmax$ in MC@NLO, and $\Qres$ in analytical resummation.
Therefore, given the fact that $\bbH$ production in the 4FS is a chief 
example for the condition of eq.~(\ref{Qshrange}) {\em not} to be fulfilled,
and that for this process one tends to use small values of $\Qres$,
it is interesting to investigate the sensitivity of NLO+PS predictions
to the choices of the parameters that appear in eq.~(\ref{murange}).
Thanks to the redundancy of the latter, we shall limit ourselves here
to studying the dependence on $\alpha$, by setting $\alpha=1,1/2,1/4$.
We consider an observable which, in NLO+PS computations, is 
maximally sensitive to the matrix-element vs parton-shower interplay,
namely the transverse momentum of the Born-level ``system'' ($\ptsyst$).
The latter is unambiguosly defined only in fixed-order calculation,
where its four momentum is the sum of the four-momenta of the Higgs,
the $b$, and the $\bb$ quark. In the case of NLO+PS simulations, we
use the sum of the four-momenta of the Higgs and of the two hardest
$B$ hadrons; no final-state cuts are applied.
\begin{figure}[h]
  \begin{center}
    \epsfig{figure=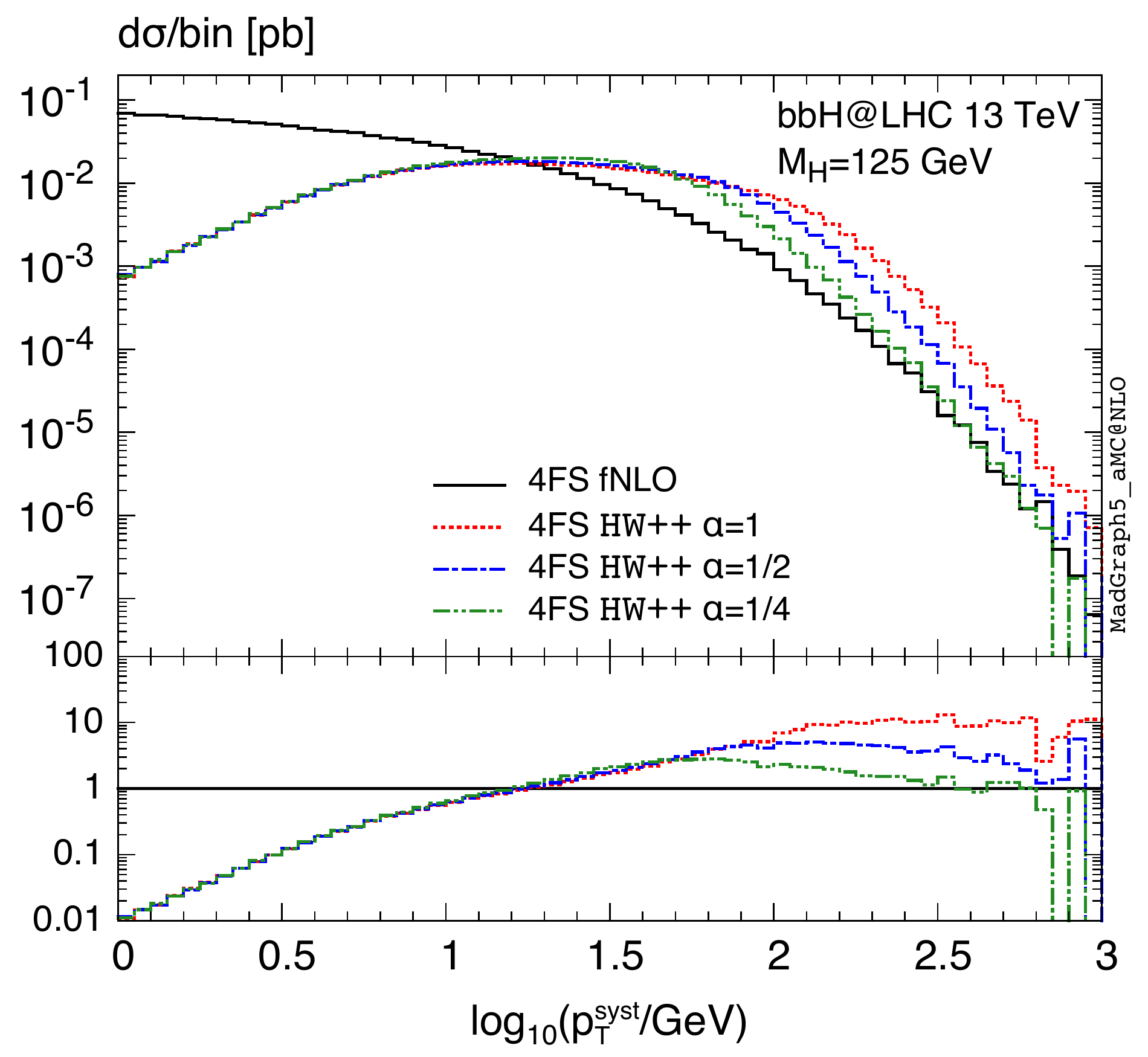,width=0.48\textwidth}
    \epsfig{figure=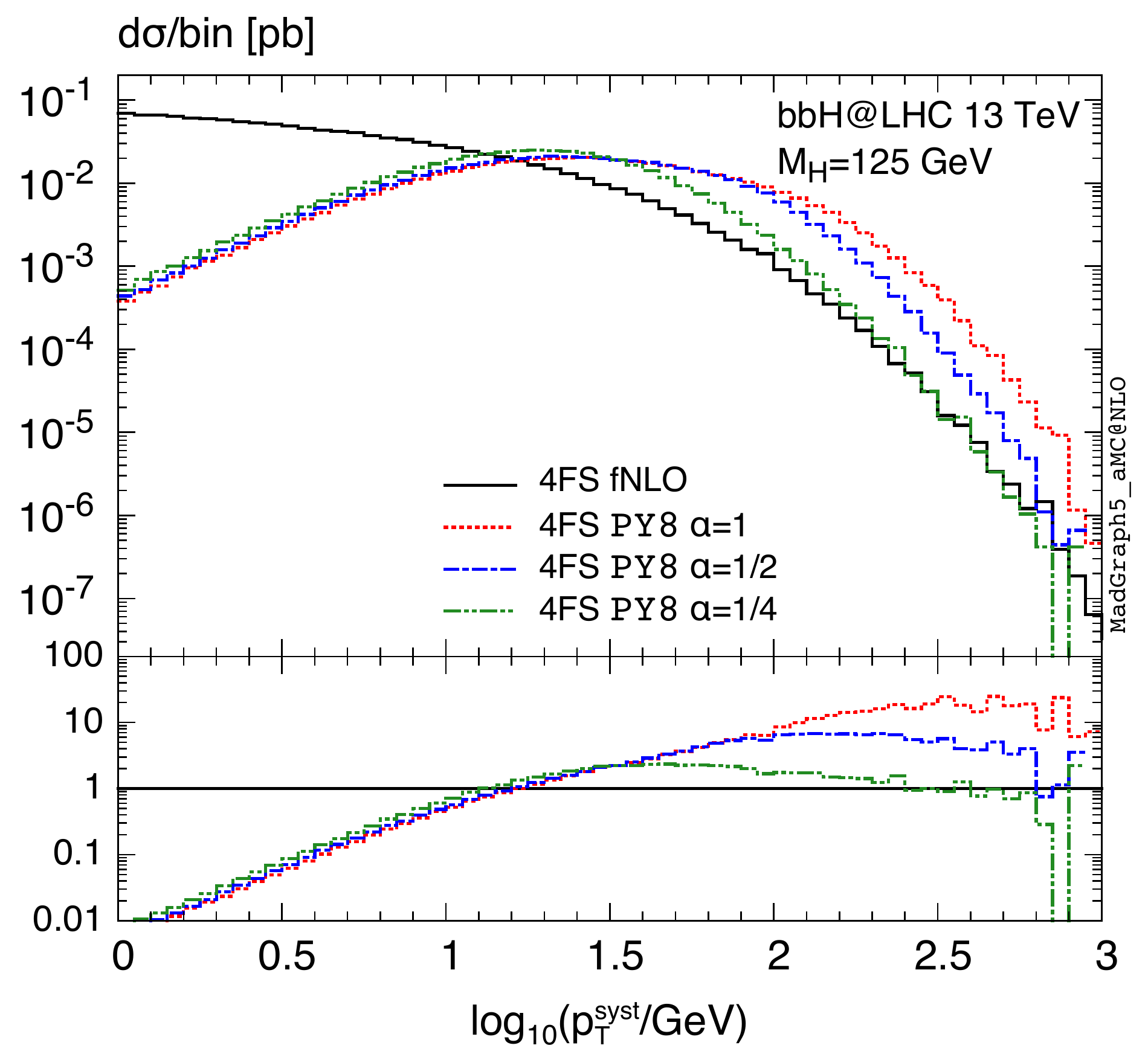,width=0.48\textwidth}
\caption{\label{fig:bbhsystem} 
Transverse momentum of the $\bbH$ or $BBH$ system, in the 4FS
at fNLO (black solid), and at NLO+PS with $\alpha=1$ (red dotted), 
$\alpha=1/2$ (blue dot-dashed), and $\alpha=1/4$ (green 
dash-double-dotted). Left panel: \HWpp; right panel: \PYe.
}
\end{center}
\end{figure}

The results are presented in fig.~\ref{fig:bbhsystem}, for both
\HWpp\ (left panel) and \PYe\ (right panel); the fNLO prediction
is shown as well. The insets display the ratio of the NLO+PS results
over the fNLO one. Only the $\yb^2$ terms ($\sigma_{\yb^2}$ in
eq.~(\ref{sig4FS})) are considered here. We remind the reader
that at sufficiently large transverse momentum the NLO+PS results 
(obtained with the MC@NLO method) will coincide by construction
with the fNLO result (in the case of $\ptsyst$, up to small effects
due to the fact that the hadron- and parton-level observables are not
exactly the same). The main message one derives from fig.~\ref{fig:bbhsystem}
is that shower (i.e.~resummation) effects, measured by the distance
between the NLO+PS and fNLO predictions, extend much farther than one
would naively expect if ``large'' values of $\alpha$ are chosen. Furthermore,
the dependence on $\alpha$ at large $\ptsyst$ is extremely significant
which, as explained above, is exceedingly rare for MC@NLO results.
On the other hand, the NLO+PS curves do behave as expected: their shapes 
show no dependence on $\alpha$ at small $\ptsyst$, and their total integrals
are equal at the level of the statistical integration error (i.e.~0.5\%), 
and equal to the fNLO rate. These observations apply to both
\HWpp\ and \PYe, which follow a rather similar pattern.
What one sees, thus, is an extremely
large matching systematics in certain corners of the phase space.
On the one hand, this effect is enhanced by the fact that the
tail of the $\ptsyst$ distribution is extremely steep. On the other hand, the 
dynamics of the process is such that the condition of eq.~(\ref{Qshrange})
appears to be significantly violated. In particular, with $\alpha=1$ the
distribution of $\Qshow$ peaks at around 180~GeV; this value decreases
to 90~GeV and 45~GeV when setting $\alpha=1/2$ and $\alpha=1/4$ respectively.
Therefore, with $\alpha=1/4$ one induces shower scales which are closer
to the values taken by the hard scales, defined as in eq.~(\ref{scref}).

At this point, there are two things which must be stressed.
Firstly, by choosing $\alpha=1$ and $\mu$ as in eq.~(\ref{scref})
one does not really introduce large logarithms in the computation.
Rather, if the ``small'' value of $\mu$ is dictated by arguments
of collinear dominance, the same arguments appear to suggest
that, since the MCs are based on a collinear approximation, small
values of $\Qshow$ have to be preferred. If these values are natural
for $\bbH$ production, by using them one makes sure that the MCs radiate
mostly away from the hard regions. An indirect confirmation of this
can be seen in fig.~\ref{fig:bbhsystem}: the smaller $\alpha$, the
closer \HWpp\ and \PYe\ are to each other. Secondly, although from
what has been said above a coherent picture emerges, one has to bear
in mind that the entire discussion stems from the very large theoretical
uncertainties that affect $\bbH$ production even at the NLO. So while
this justifies, to a certain extent, the practice of finding optimal
scale choices, it does not allow one to ignore the existence of large
systematics, which might prove to be crucial in the comparisons with data.

Owing to the arguments above, $\alpha=1/4$ is our default choice
in 4FS (N)LO+PS simulations, whose results we are now going to present.
It is interesting, and reassuring for the self-consistency of the
theoretical description of $\bbH$ production, that the hard scales
relevant to our novel NLO+PS calculation appear to follow the same
pattern as those relevant to other approaches to $\bbH$ production.
In this paper, we shall refrain from quoting an uncertainty associated
with the variation of $\alpha$.

We start by reporting, in table~\ref{tab:rates}, the results for
the total cross sections, both fully inclusive and within cuts.
As far as the latter are concerned, we have considered three
possibilities: the requirement that there be at least one or two
$b$-jet(s) (see sect.~\ref{sec:input} for the jet-finding parameters),
and that the transverse momentum of the Higgs be larger than 
100~GeV (boosted scenario). In the case of the fully inclusive
cross section, we have also computed it by setting $\muR=\muF=(\mH+2\mb)/4$,
for ease of comparison with the results in the literature, with
which we find agreement at the level of the numerical integration
errors. For the cross sections within cuts,
since they depend on the kinematics of final-state objects, we report
the results obtained both with \PYe\ and with \HWpp. The fractional
scale uncertainties that we show in table~\ref{tab:rates} are 
computed by varying the scales as indicated in eq.~(\ref{scalevar});
since they are largely MC-independent, we give them
only in the case of the \PYe\ simulations. We present separately the
results for the $\yb^2$ and $\yb\yt$ terms.
\begin{table}
\begin{center}
\begin{tabular}{cl|cccc}
\toprule
\multicolumn{2}{c|}{\multirow{2}{*}{$\sigma$[pb]}} & 
\multicolumn{3}{c}{NLO} & LO \\
& & $\yb^2$ & $\yb\yt$ & $\yb^2+\yb\yt$ & $\yb^2$ \\
\midrule
\multicolumn{2}{c|}{inclusive} & 
$0.448^{+ 19.8 \% }_{- 20.8\%}$ &
$-0.0365^{+ 35.5\%}_{-62.8\%}$ & 
$0.411^{+ 24.6\%}_{-28.4\%}$ &
$0.478^{+ 59.0\%}_{- 34.6\%}$ \\
\multicolumn{2}{c|}{inclusive $(\mu=\frac{\mH+2\mb}4)$} & 
$0.515$ &
$-0.0430$ &
$0.472$ &  
$0.540$ \\
\multirow{2}{*}{$\ge 1j_b$} &
\PYe\ &
$0.133^{+16.7 \% }_{- 17.3 \% }$ &
$-0.0148^{+35.0\%}_{- 60.1\%}$ &
$0.118^{+23.5\%}_{- 26.8\%}$ &
$0.150^{+ 55.9\%}_{- 32.8\%}$ \\
 & 
\HWpp\ &
$0.119$ &
$-0.0123$ &
$0.107$ &
$0.120$ \\
\multirow{2}{*}{$\ge 2j_b$} &
\PYe\ &
$0.0133^{+ 13.7 \% }_{- 16.0 \% }$ &
$-0.00147^{+ 34.3\%}_{- 58.8\%}$ &
$0.0118^{+ 20.0\%}_{- 25.1\%}$ &
$0.0168^{+54.4\%}_{- 32.7\%}$ \\
 &
\HWpp\ &
$0.0121$ &
$-0.000955$ &
$0.0112$ &
$0.0120$ \\
\multirow{2}{*}{\scriptsize $\pt(H)\ge 100$~GeV} &
\PYe\ &
$0.0123^{+ 21.9 \% }_{- 20.0 \% }$ &
$-0.00167^{+ 34.6\%}_{- 58.9\%}$ &
$0.0106^{+ 30.7\%}_{- 32.5\%}$ &
$0.0117^{+ 57.6\%}_{- 33.7\%}$ \\
 &
\HWpp\ &
$0.0122$ &
$-0.00144$ &
$0.0107$ &
$0.0106$ \\
\bottomrule
\end{tabular}
\end{center}
\caption{\label{tab:rates}
Predictions for the total rates (in pb) in the 4FS. 
See sect.~\ref{sec:input} for the choices of input parameters.
}
\end{table}
The conclusions that can be drawn from the table are the following.
\begin{enumerate}
\item The inclusion of NLO corrections reduces in a very significant manner
the scale dependence of the $\yb^2$ terms w.r.t.~that at the LO;
the residual scale dependence is however still large ($\sim\! 20\%$).
\item The effect on the fully-inclusive $\yb^2$ cross section of the NLO
corrections is moderate ($\sim\! 10\%$) and negative (i.e.~the $K$ factor is 
slightly smaller than one).
\item There are large differences ($\sim\! 15\%$) between the cross sections 
computed with dynamic or fixed hard scales, of the same order as the NLO 
scale uncertainties.
\item The contribution of the $\yb\yt$ term is negative and non negligible in the 
\sm{} ($\sim\! 10\%$), although within the scale uncertainties of the
$\yb^2$ NLO results.\footnote{
The relative size of the $\ybyt$ term is determined by the coupling
structure in the relevant theory; for example, in 2HDM models such a term
will become negligible at large $\tan\beta$.}  Its scale dependence is larger than that of
the $\yb^2$ term (as expected, owing to the absence of a $\yb\yt$ 
contribution at the LO).
\item In the cases of the cross sections within cuts, the inclusion of
NLO corrections improves the agreement between the predictions of the
two MCs w.r.t.~that at the LO.
\item The effects of the cuts are significant: the cross section is
reduced by a factor larger than three when requiring at least one $b$-jet,
and by a further factor of about ten when a second $b$-jet is tagged.
The boosted-Higgs case is similar to the two-jet one.
\end{enumerate}

\noindent
We now turn to discussing differential observables. Unless stated otherwise, 
we shall limit ourselves to presenting the results relevant to the $\yb^2$
terms since, apart from these being dominant, it turns out that the 
$\yb\yt$ contribution is fairly flat for most observables; there is
one striking exception, which we shall discuss in details later.
The figures of this section and of sect.~\ref{sec:4FSvs5FS} are
generally organised according to the following pattern. There is a main
frame, where the relevant predictions (e.g.~NLO+PS, fNLO, and so forth) 
are shown with their absolute normalisation, and as cross section per
bin (namely, the sum of the contents of the bins is equal to the total
cross section, possibly within cuts). In an inset we display the bin-by-bin
ratio of all the histograms which appear in the main frame over one of them,
chosen as a reference. Finally, in a second inset the bands that represent
the fractional scale dependence are given: they are computed by taking the
bin-by-bin ratios of the maximum and the minimum (obtained according to
eq.~(\ref{scalevar})) of a given simulation over the same central
prediction that has been used as a reference for the ratios of 
the first inset. In this paper we have ignored the PDF systematics;
however, we stress that it can be included at no extra CPU cost, using
the same reweighting procedure~\cite{Frederix:2011ss} employed here
for the scale uncertainty.

\begin{figure}[h]
  \begin{center}
    \epsfig{figure=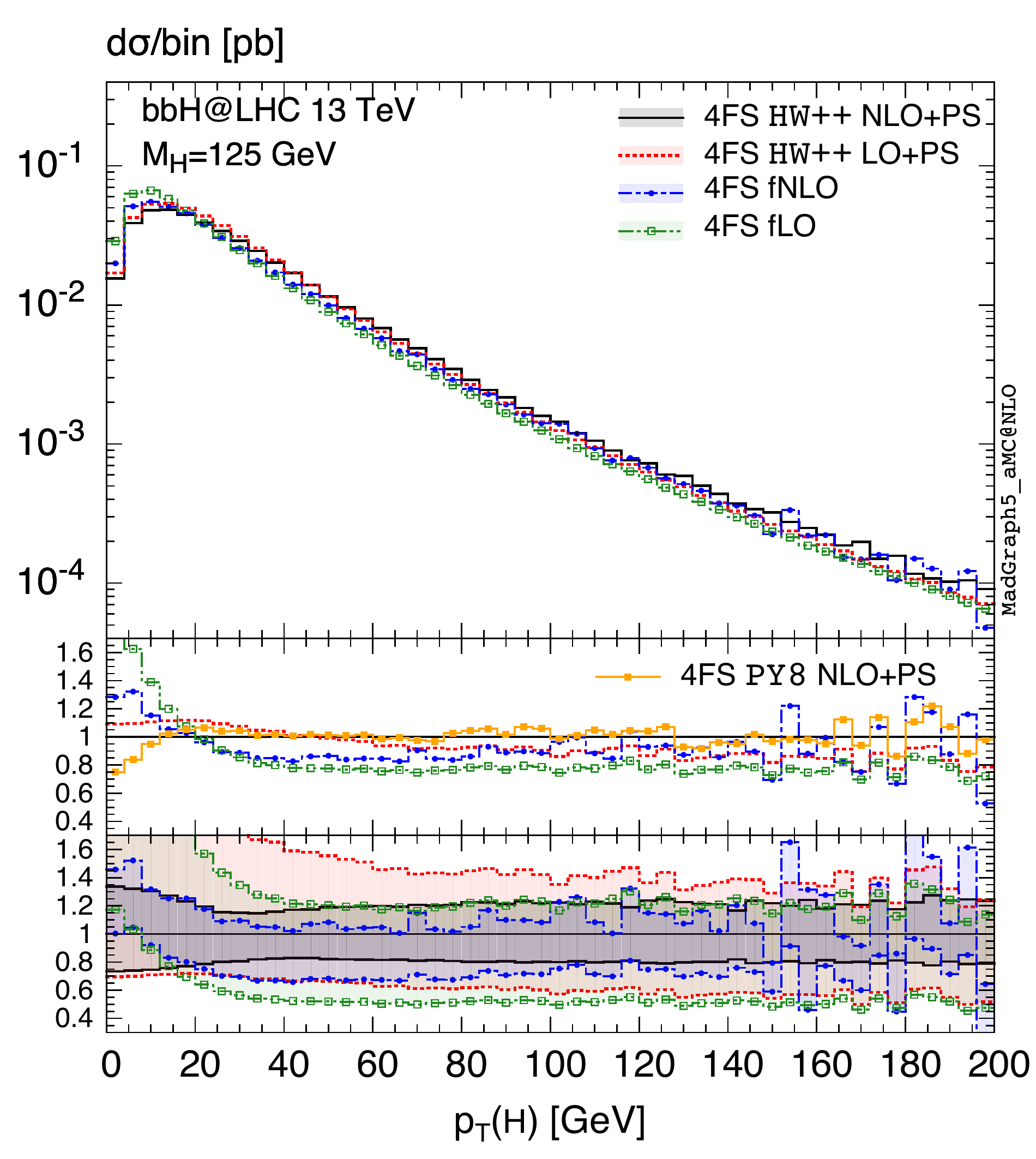,width=0.48\textwidth}
    \epsfig{figure=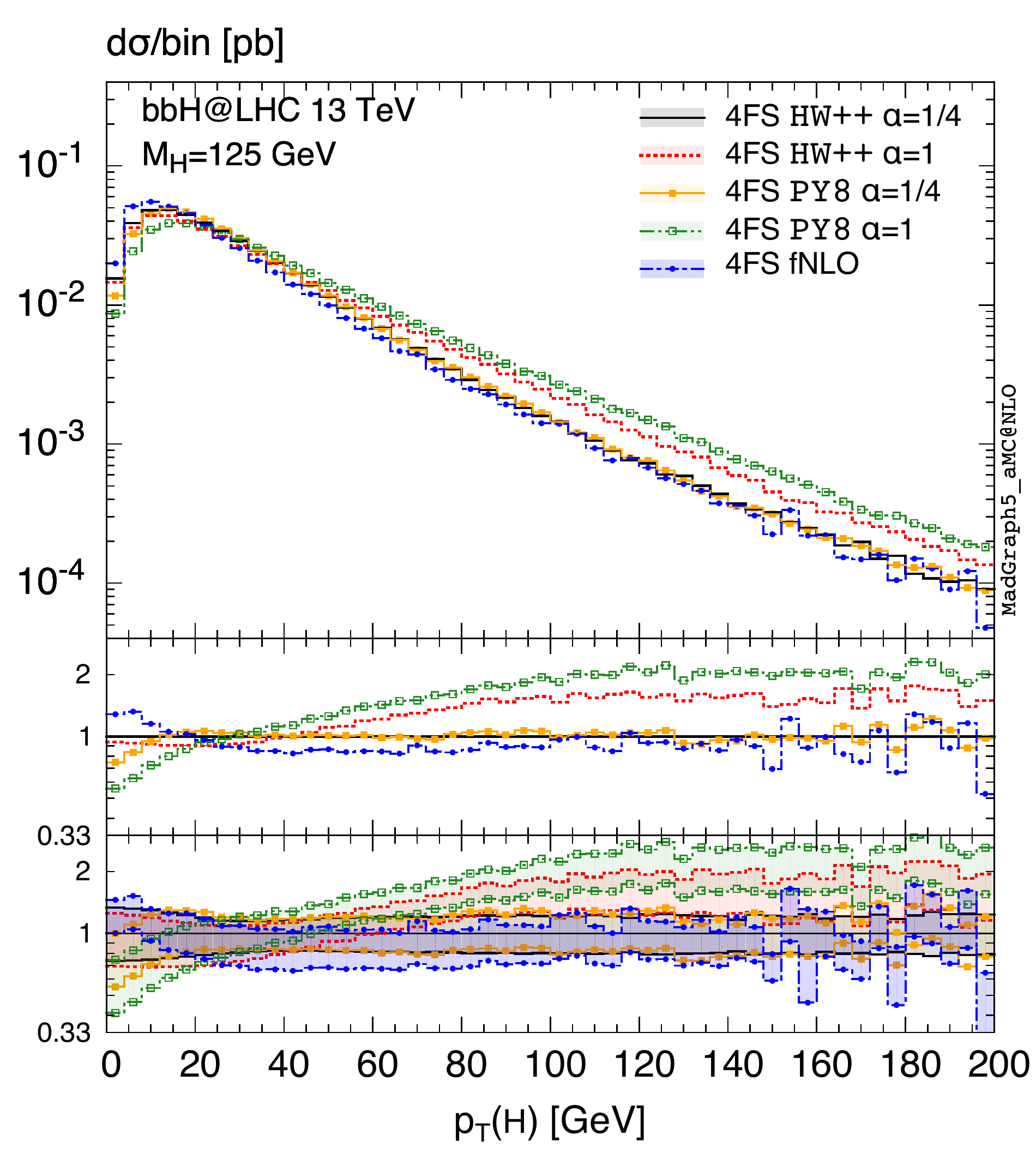,width=0.48\textwidth}
\caption{\label{fig:pTH4FS} 
Higgs transverse momentum. See the text for details.
}
\end{center}
\end{figure}
We begin in fig.~\ref{fig:pTH4FS} with the transverse momentum of the Higgs.
In the left panel we present the NLO+PS \HWpp\ result (black solid; this 
is the reference histogram), its LO+PS counterpart (red dotted), and
the fNLO (blue dot-dashed) and fLO (green dash-double-dotted) predictions.
In the upper inset, on top of the ratios of the curves that appear in
the main frame we also display the ratio of the NLO+PS \PYe\ result
(orange solid with full boxes) over the \HWpp\ one. In the region
$\pt(H)\gtrsim 20$~GeV, all results are within 25\% of each other.
The agreement between the NLO+PS \HWpp\ and \PYe\ predictions is
excellent (they are essentially identical up to statistics); the
fNLO result is slightly harder than its showered counterparts, and
it gets closer to them with increasing $\pt(H)$. The LO+PS \HWpp\
prediction is within 10\% of the NLO+PS one in the whole $\pt(H)$
range. \PYe\ differs visibly from \HWpp\ at the NLO+PS
only in the small $\pt(H)$ region, where the difference is however
20\% at most; on the other hand, in that region it is the fixed-order
results which display the largest discrepancies w.r.t.~our \HWpp\
NLO+PS reference curve. From the lower inset one sees the very significant
reduction of the scale dependence at the NLO w.r.t.~the one at the LO,
analogous to that already observed in table~\ref{tab:rates} for total 
cross sections. The uncertainty bands of the NLO predictions almost 
completely overlap, and they are by and large contained within 
those relevant at the LO.

The \HWpp\ and \PYe\ NLO+PS results, and the fNLO one, are also shown
in the right panel of fig.~\ref{fig:pTH4FS}, with the same patterns as
those employed in the left panel of that figure; on top of these, we also 
present the \HWpp\ (red dotted) and \PYe\ (green dash-double-dotted with
open boxes) NLO+PS predictions obtained with $\alpha=1$. Although the 
effects due to the change of $\alpha$ are not as large as those affecting 
$\ptsyst$ (see fig.~\ref{fig:bbhsystem}), they are very significant in
the large-$\pt(H)$ region, where the two NLO+PS results obtained with
the same MC have non-overlapping scale-uncertainty bands (the more so
for \PYe). Furthermore, while when $\alpha=1/4$ \HWpp\ and \PYe\ agree
perfectly with each other, this is not the case when $\alpha=1$ (although
their uncertainty bands do overlap). Note that, as expected, the value
of $\alpha$ does not affect the shape of the small-$\pt(H)$ distribution
for a given MC. Overall, these results confirm the findings discussed
above for $\ptsyst$, and the fact that $\alpha=1/4$ is a sensible choice.

\begin{figure}[t]
  \begin{center}
    \epsfig{figure=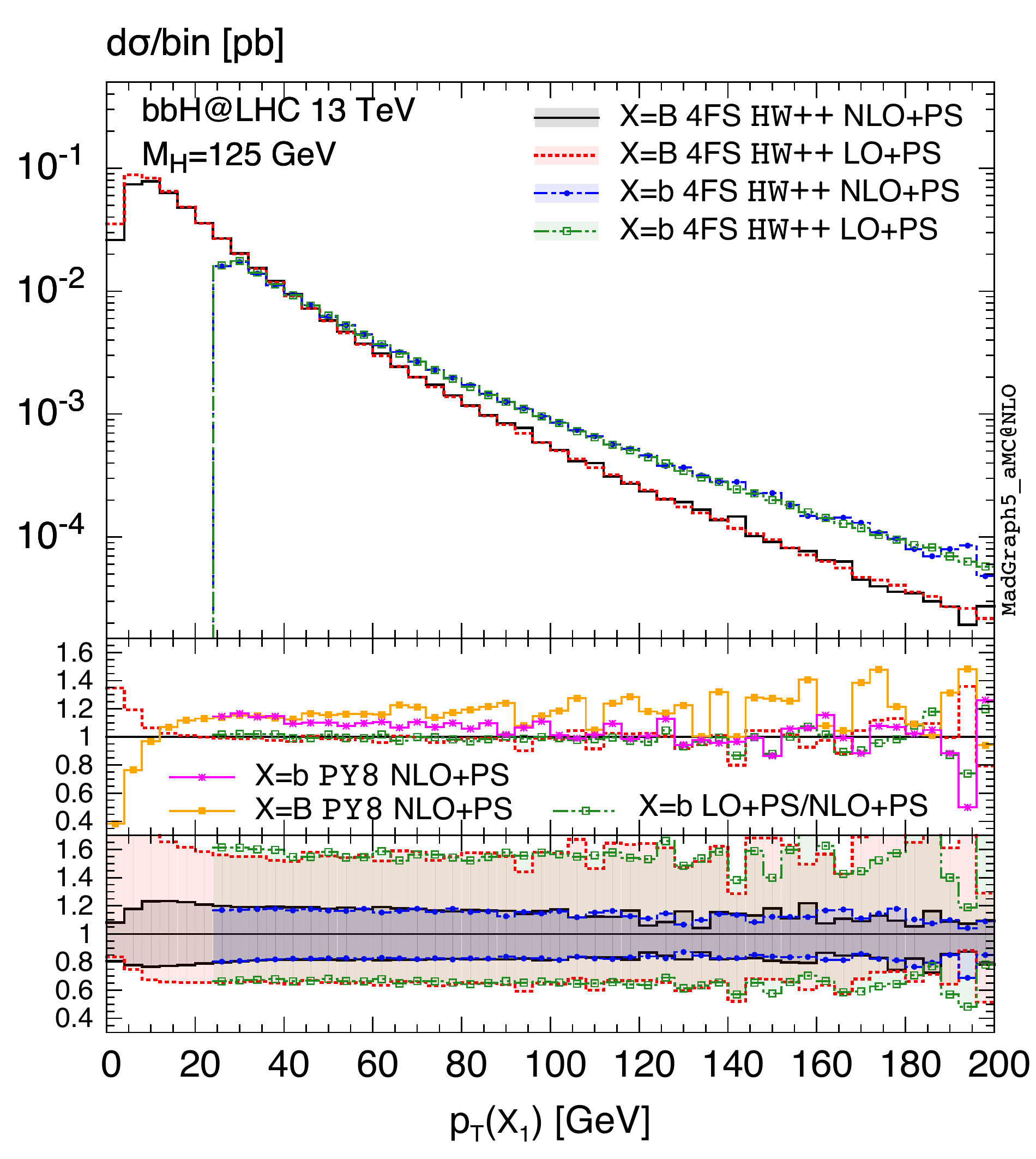,width=0.48\textwidth}
    \epsfig{figure=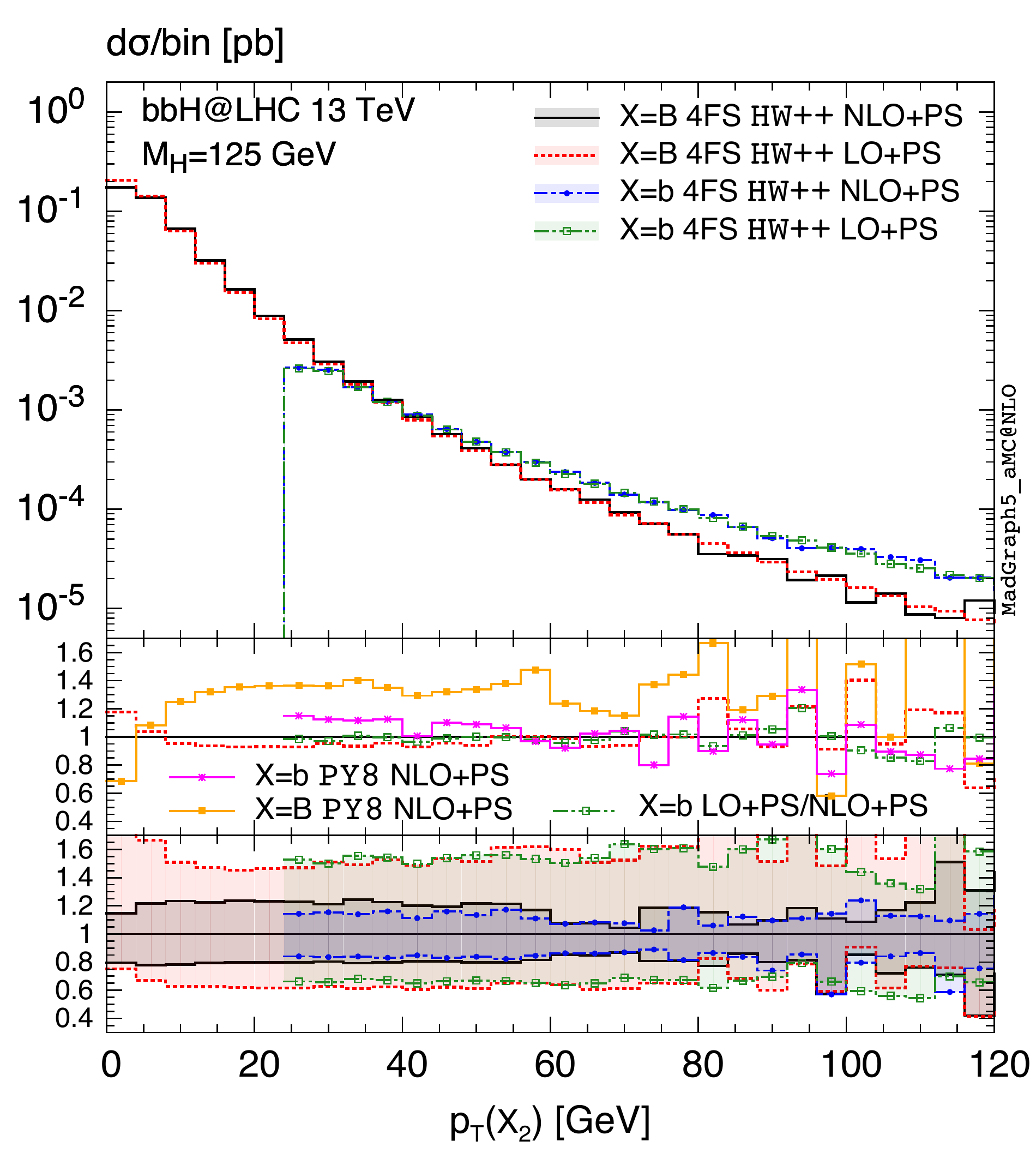,width=0.48\textwidth}
\caption{\label{fig:pTX1} \HWpp\ predictions for the transverse
momentum of hardest (left panel) and second-hardest (right panel)
$B$ hadron and $b$-jet. See the text for details.
}
\end{center}
\end{figure}
\begin{figure}[t]
  \begin{center}
    \epsfig{figure=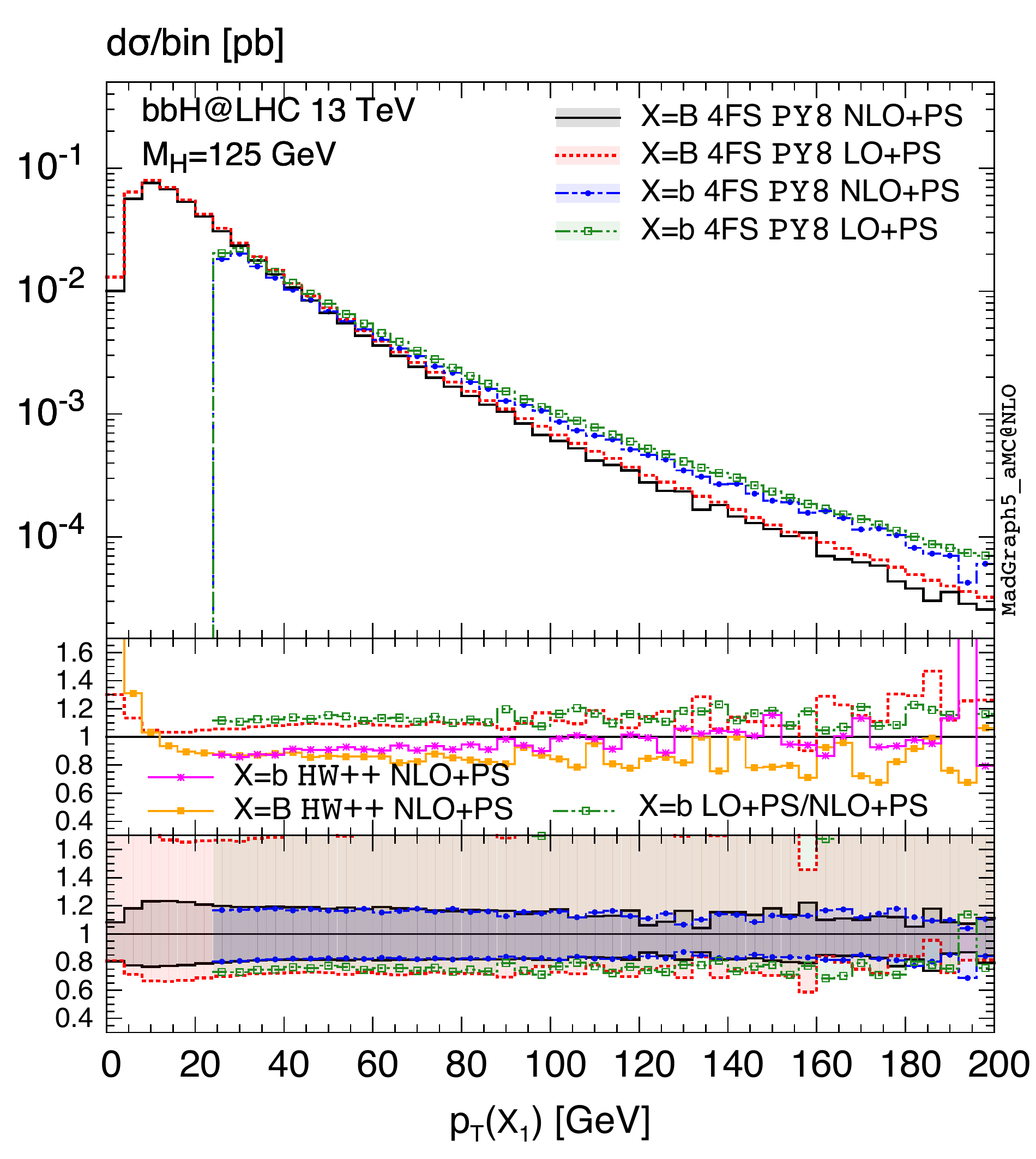,width=0.48\textwidth}
    \epsfig{figure=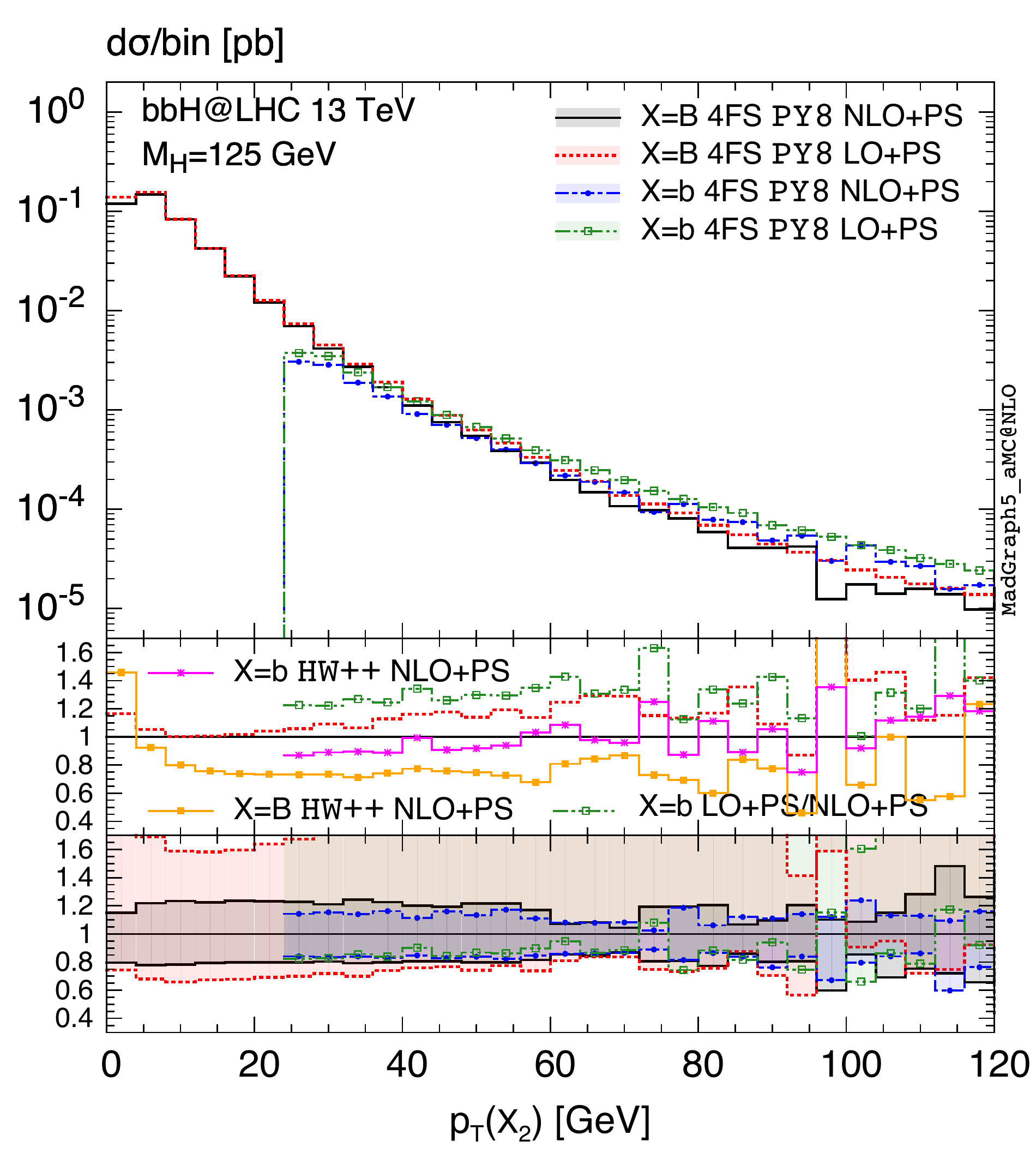,width=0.48\textwidth}
\caption{\label{fig:pTX1py8}
Same as in fig.~\ref{fig:pTX1}, but for \PYe.
}
\end{center}
\end{figure}

We next consider the transverse momentum distributions of quantities related
to single $b$ quarks: the hardest $B$ hadron and $b$-jet, shown in the left 
panels of fig.~\ref{fig:pTX1} (for \HWpp) and of fig.~\ref{fig:pTX1py8}
(for \PYe) at the NLO+PS and LO+PS accuracies, and the second-hardest
$B$ hadron and $b$-jet, shown in the right panels of the same figures.
The results for $B$ hadrons are presented as black solid (NLO+PS) and 
red dotted (LO+PS) histograms, while those for $b$-jets are the
blue dot-dashed (NLO+PS) and green dash-double-dotted (LO+PS) histograms.
In the insets, the red dotted (green dash-double-dotted) histograms display
the ratios of the LO+PS over the NLO+PS predictions for $B$ hadrons ($b$-jets).
In the case of \HWpp, the inclusion of the NLO corrections is hardly 
visible, except at the threshold of the $B$-hadron distributions, where
the differences account for the smaller-than-one $K$ factor reported
in table~\ref{tab:rates} for total cross sections (the NLO+PS results being
smaller than the LO+PS ones). The $b$-jet distributions are
essentially identical, in shapes and rates, at the NLO+PS as at the LO+PS.
In the case of \PYe, one observes similar effects as for \HWpp\ at the 
thresholds of the $B$-hadron distributions, but also a softening of 
the high-$\pt$ spectra at the NLO+PS w.r.t.~the LO+PS, softening which is
more pronounced in the case of the second-hardest $B$ hadron. As far
as $b$-jets are concerned, the NLO+PS and LO+PS results are quite similar 
in shape, but the rates of the former are about 10\% and 20\% lower than 
those of the latter for the hardest and second-hardest jet,
respectively. Still, for each observable (a fortiori for \HWpp) 
the NLO+PS uncertainty band lies completely within that of the corresponding
LO+PS prediction. The comparison of the two MCs for a given observable
is best read from the upper insets, where we report the ratios \PYe/\HWpp\
(fig.~\ref{fig:pTX1}) and  \HWpp/\PYe\ (fig.~\ref{fig:pTX1py8})
as solid orange (overlayed with boxes) and solid magenta (overlayed
with stars) histograms for $B$ hadrons and $b$-jets, respectively; we 
limit ourselves to presenting these ratios at the NLO+PS accuracy. For 
$\pt$'s larger than about 20~GeV \HWpp\ and \PYe\ agree quite well with 
each other in terms of shapes; rate-wise, \PYe's are larger than \HWpp's, 
with the largest differences (about 30\%) occurring for the second-hardest
$B$-hadron, and the smallest (about 10\%) for $b$-jets. Significant
differences (i.e.~outside of the uncertainty bands) are seen only in
the $\pt\sim 0$ region for $B$ hadrons, where \PYe\ can be up to
a factor of two smaller than \HWpp\ in the first bin of the 
hardest-$B$ hadron distribution.

The vast majority of the differential observables which we have investigated
follow the pattern that underpins figs.~\ref{fig:pTH4FS}--\ref{fig:pTX1py8}:
the various approximations are consistent within their associated
theoretical uncertainties, and there are small corrections due to NLO 
and/or parton-shower effects and an overall agreement between the \HWpp\
and \PYe\ predictions. A couple of exceptions are however notable,
which we are now going to address explicitly.

\begin{figure}[t]
  \begin{center}
    \epsfig{figure=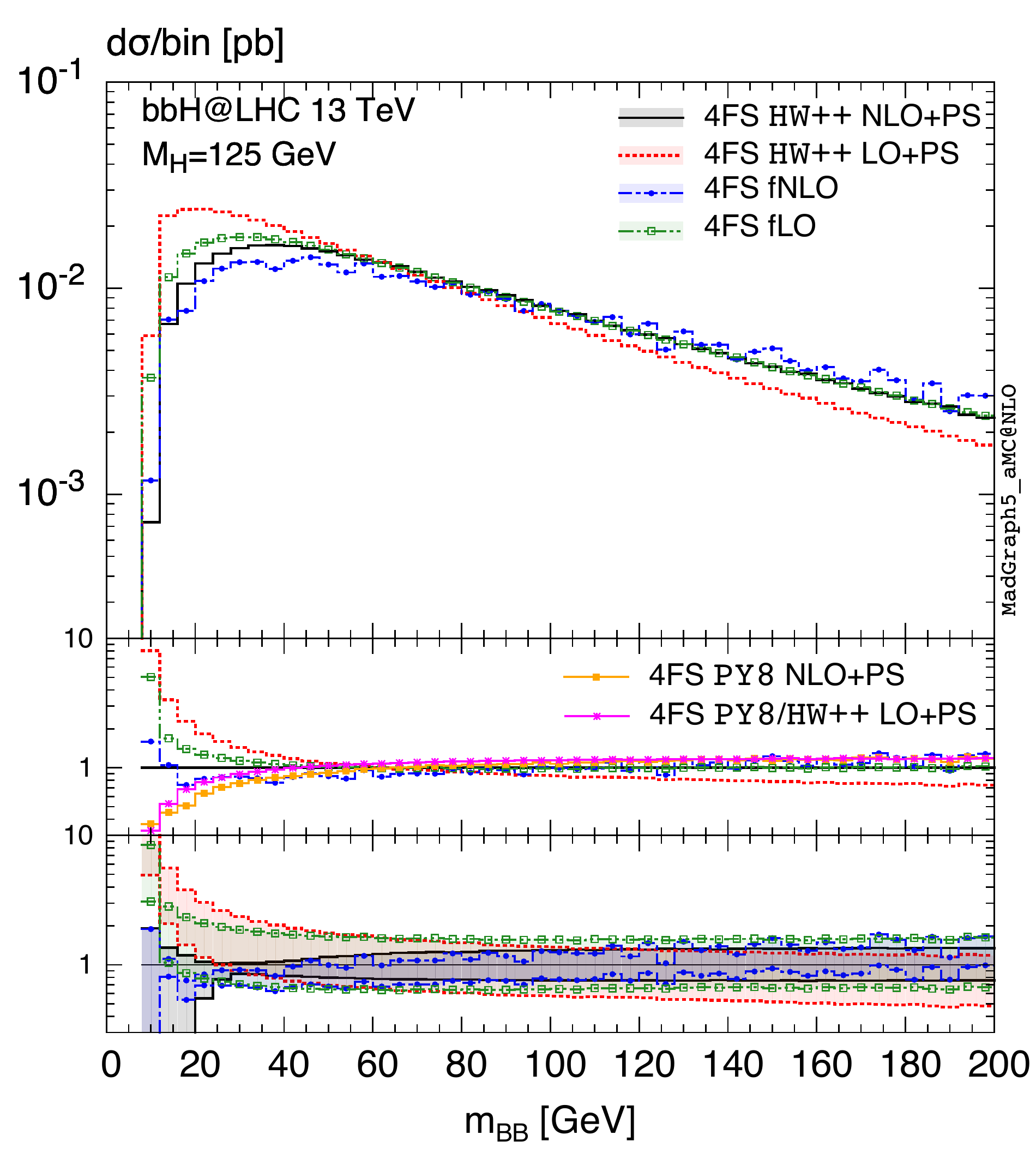,width=0.48\textwidth}
    \epsfig{figure=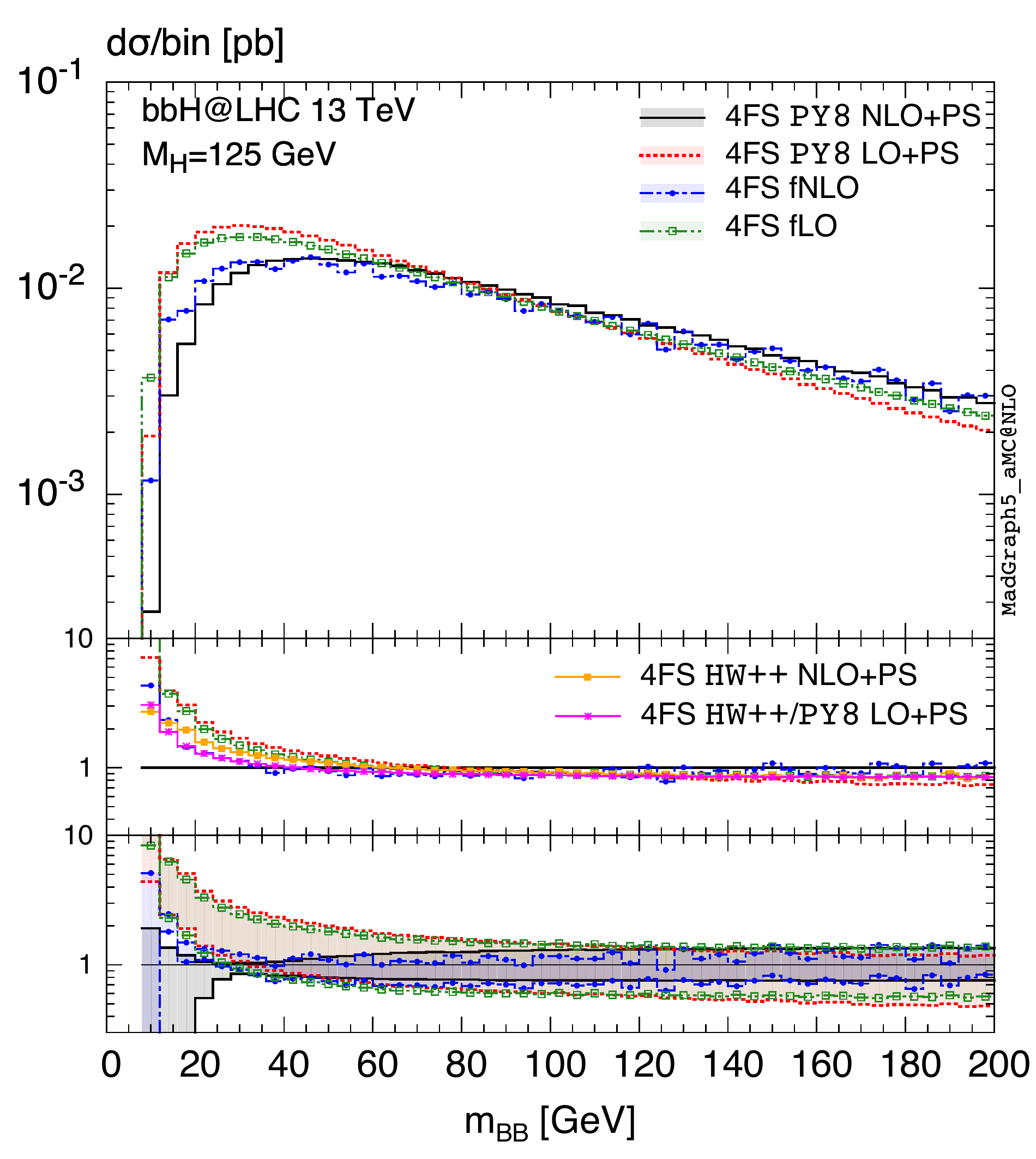,width=0.48\textwidth}
\caption{\label{fig:4FSmBB} 
Invariant mass of the two hardest $B$ hadrons (at (N)LO+PS)
or two $b$ quarks (at f(N)LO). Left panel: \HWpp; right panel: \PYe.
See the text for details.
}
\end{center}
\end{figure}
In fig.~\ref{fig:4FSmBB} we present various predictions for the
invariant mass of the two hardest $B$ hadrons at the NLO+PS
(black solid) and LO+PS (red dotted) accuracy, and that of the two
$b$ quarks at the fNLO (blue dot-dashed) and fLO (green dash-double-dotted)
accuracy; no cuts are applied, and both quantities have been labelled
$m_{BB}$ for simplicity. The left and right panels display
the \HWpp\ and \PYe\ results respectively. In the upper insets,
on top of showing the ratios of the histograms that appear in the
main frames over the relevant reference curves (\HWpp\ NLO+PS 
and \PYe\ NLO+PS in the left and right panels respectively),
we also show the ratios \PYe/\HWpp\ (left panel) and \HWpp/\PYe\
(right panel) at the NLO+PS (solid orange overlayed with boxes) and 
at the LO+PS (solid magenta overlayed with stars) accuracy.
The most obvious feature of fig.~\ref{fig:4FSmBB} is the behaviour
at threshold, where the differences between the two MCs are extremely
large, and to a good extent independent of the perturbative order
considered. The LO+PS \HWpp\ prediction is quite far from the fLO
result, which in turn is in reasonable agreement with \PYe's.
One should be careful not to interpret this fact as an evidence
for the \PYe\ curve to be more realistic: the threshold region is
where one expects to be more sensitive to the details of the $b\to B$
hadronisation, and thus where the differences between the invariant 
mass of the $B$ hadrons and that of the $b$ quarks are largest.
The inclusion of NLO effects, regardless of the presence of parton
showers, hardens the spectra in a very significant manner; owing
to the shape of the LO+PS \HWpp\ result, the shift in the peak position
when going from LO to NLO is particularly large in the case of that
MC. A remarkable fact is the similarity of the pattern 
LO+PS$\to$NLO+PS in \PYe\ and \HWpp, in spite
of the differences of the individual results at a given order: this
can be seen clearly from the red dotted histograms in the upper insets
of the two panels. Away from the threshold region, the two MCs are 
quite close to each other (and close to the fixed-order result, too)
at the NLO; the agreement is worse, but still amply within theoretical
uncertainties, at the LO, where however the fLO prediction is clearly
farther away from the \HWpp's than from the \PYe's. Overall, \PYe\
results are harder than those of \HWpp, but the difference in hardness 
descreases with the perturbative order. We finally point out that the 
differences in the threshold region we have discussed above are greatly 
reduced or absent when one requires the presence of at least one large-$\pt$
$b$ quark (e.g.~in the invariant mass of two $b$-jets, or when cutting
away the low-$\pt(B)$ region).

\begin{figure}[h]
  \begin{center}
    \epsfig{figure=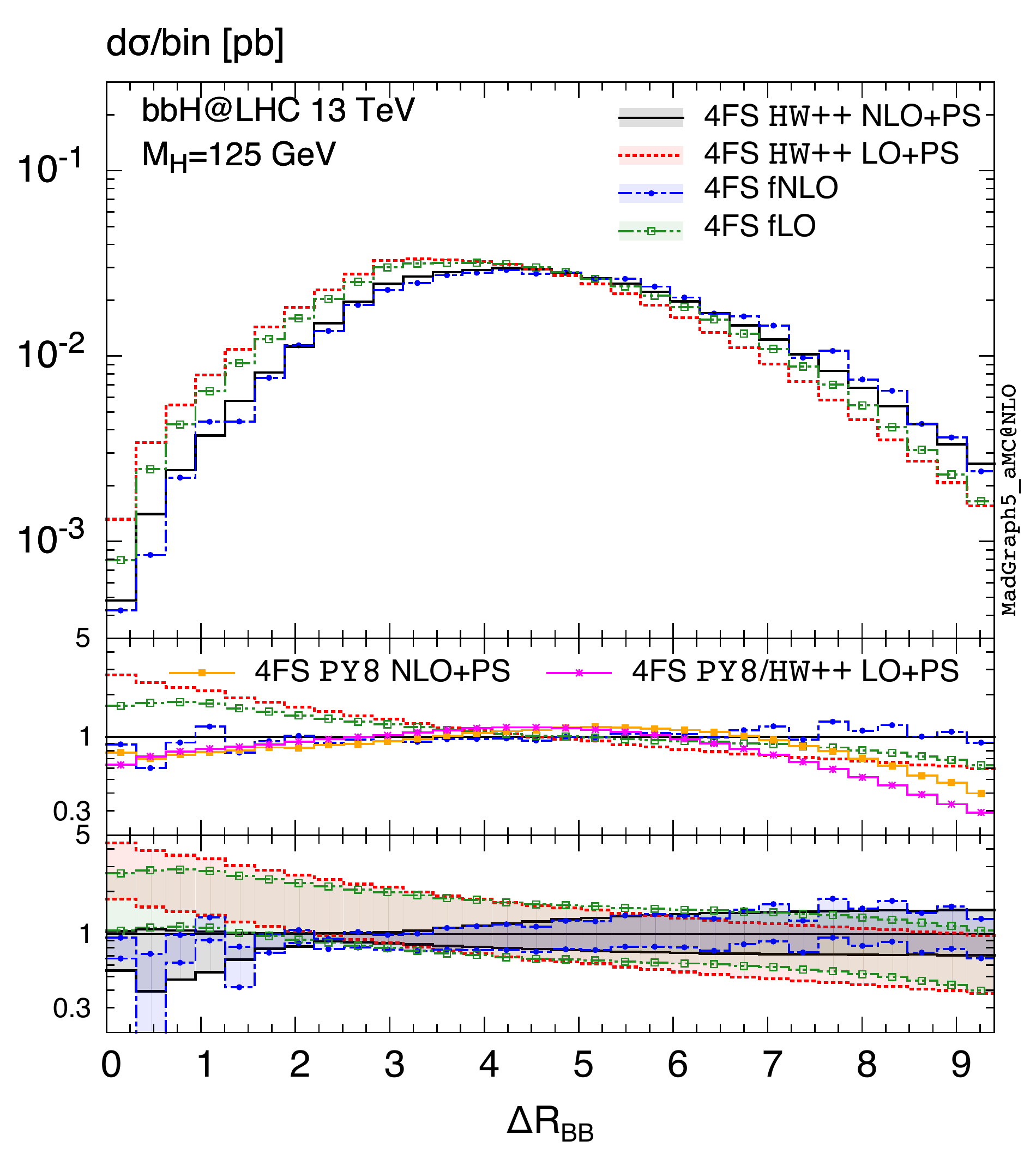,width=0.48\textwidth}
    \epsfig{figure=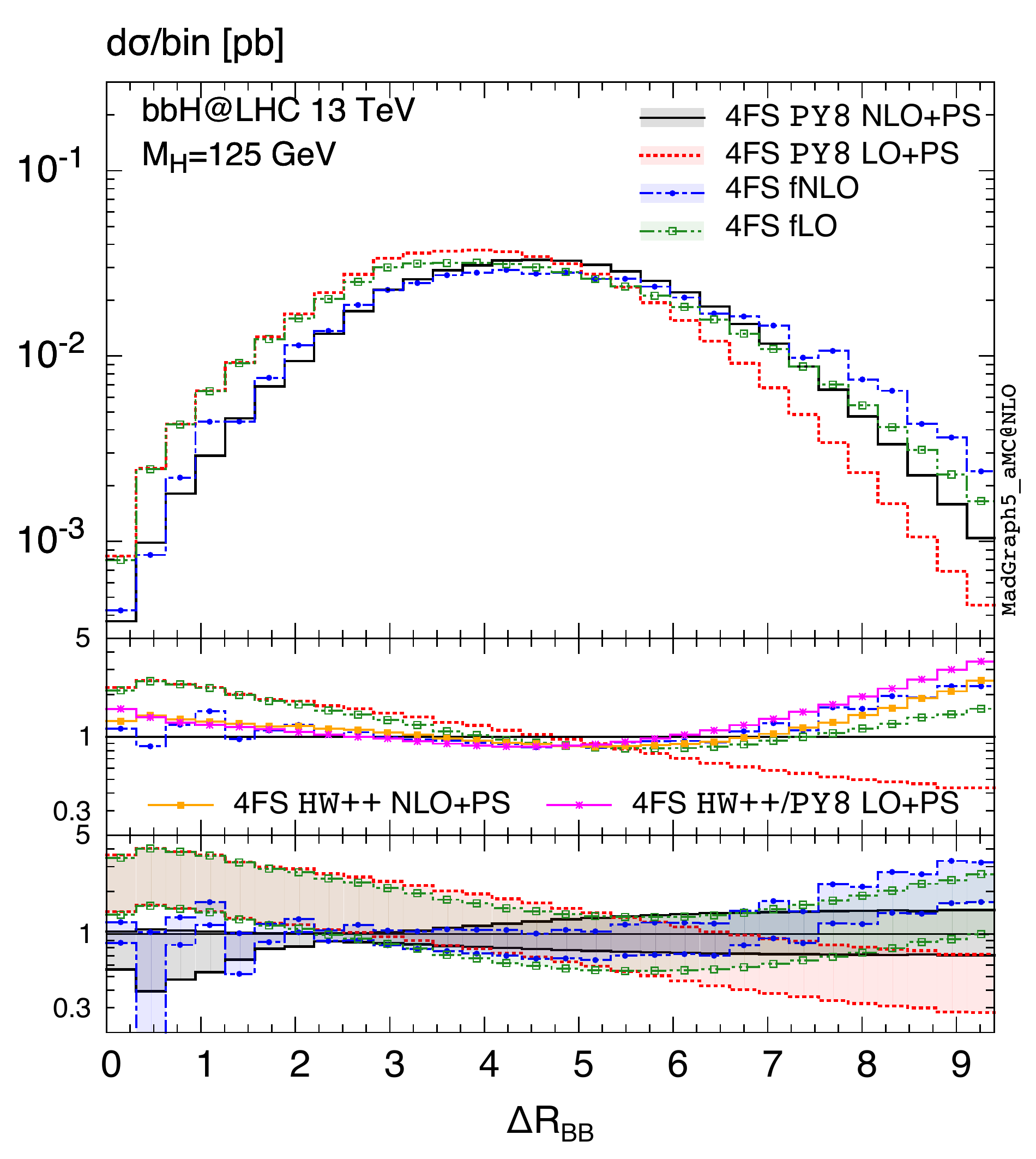,width=0.48\textwidth}
\caption{\label{fig:4FSDeltaRBB} 
Same as in fig.~\ref{fig:4FSmBB}, for the distance 
in the $\eta\!-\!\varphi$ plane.
}
\end{center}
\end{figure}
The second interesting, if less spectacular, case of significant differences
among predictions is presented in fig.~\ref{fig:4FSDeltaRBB}, where we
show the distance in the $\eta\!-\!\varphi$ plane between the two hardest 
$B$ hadrons (at (N)LO+PS) or the two $b$ quarks (at f(N)LO); both are denoted
by $\Delta R_{BB}$ for simplicity. As in the case of $m_{BB}$ previously
discussed, no cuts are applied. The layout of fig.~\ref{fig:4FSDeltaRBB}
is the same as that of fig.~\ref{fig:4FSmBB}. The salient feature
of the present results is to be found at large $\Delta R_{BB}$
(which corresponds to a large separation in pseudorapidity
between the two $B$ hadrons or $b$ quarks). In that
region, both MCs tend to be lower than the corresponding fixed-order
result; however, in the case of \PYe\ the decrease of the cross
section is dramatic (even at the NLO, where it is nonetheless smaller than 
at the LO), while it is modest (and within scale uncertainty) for \HWpp.
Over the whole spectrum, NLO corrections have a large impact, depleting
the cross section at small $\Delta R_{BB}$ and enhancing it at
large $\Delta R_{BB}$. This pattern is independent of whether parton
showers are included; note that in some regions the LO and NLO uncertainty
bands do not overlap. Up to $\Delta R_{BB}\sim 6$ and $\Delta R_{BB}\sim 7.5$ 
(at the LO and NLO, respectively), the mutual agreement between MCs and with
fixed-order results is quite good, but it rapidly degrades above those 
values as was already discussed; for all $\Delta R_{BB}$, however, the
agreement at the NLO is better than that at the LO. We finally remark
that, in terms of the comparison of the MC predictions with the fixed-order
ones, the situation at large $\Delta R_{BB}$ is the opposite as that
for $m_{BB}$ at threshold, with \PYe\ (N)LO+PS being farther away
from f(N)LO than \HWpp\ for the former observable, and closer for the latter.
While, as was stressed before, the level of agreement with f(N)LO results
cannot be used as a discriminant for MCs, this fact underlines the different
characteristics of different MCs, and the necessity of a conservative
estimate of MC systematics.

We now turn to discussing the impact of the $\yb\yt$ term
(i.e.~the $\sigma_{\yb\yt}$ contribution to the NLO cross section
in eq.~(\ref{sig4FS})). We have investigated a rather large number
of observables, and generally found such a term to be flat at
the level of 5\%--20\% (this fraction is measured by evaluating the 
ratio of the $\yb\yt$ over the $\yb^2$ contribution, and by taking its
distance from a horizontal line); we remind the reader (see 
table~\ref{tab:rates}) that the $\yb\yt$ term gives a negative 
contribution of the order of 10\% at the level of total rates.
Therefore, although there is no reason to neglect the $\yb\yt$
contribution, an overall rescaling is a decent approximation
(also in view of the large scale uncertainties that affect the 
$\yb^2$ $b\bb H$ cross section) for most observables. There are two
counterexamples, one of which is particularly spectacular, which
we present below; perhaps not surprisingly, they coincide with
the $m_{BB}$ and $\Delta R_{BB}$ observables which we have just discussed.

\begin{figure}[t]
  \begin{center}
    \epsfig{figure=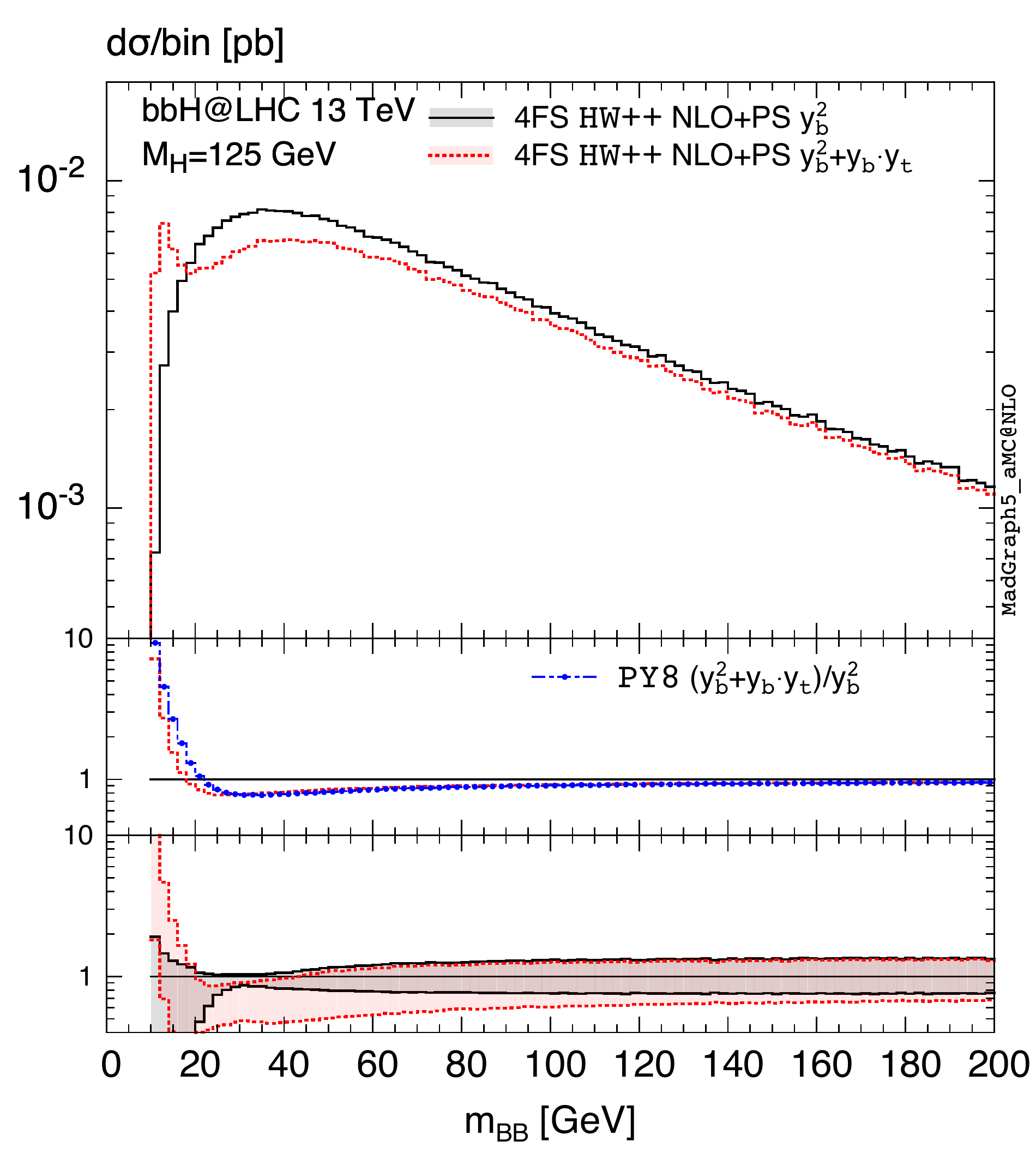,width=0.48\textwidth}
    \epsfig{figure=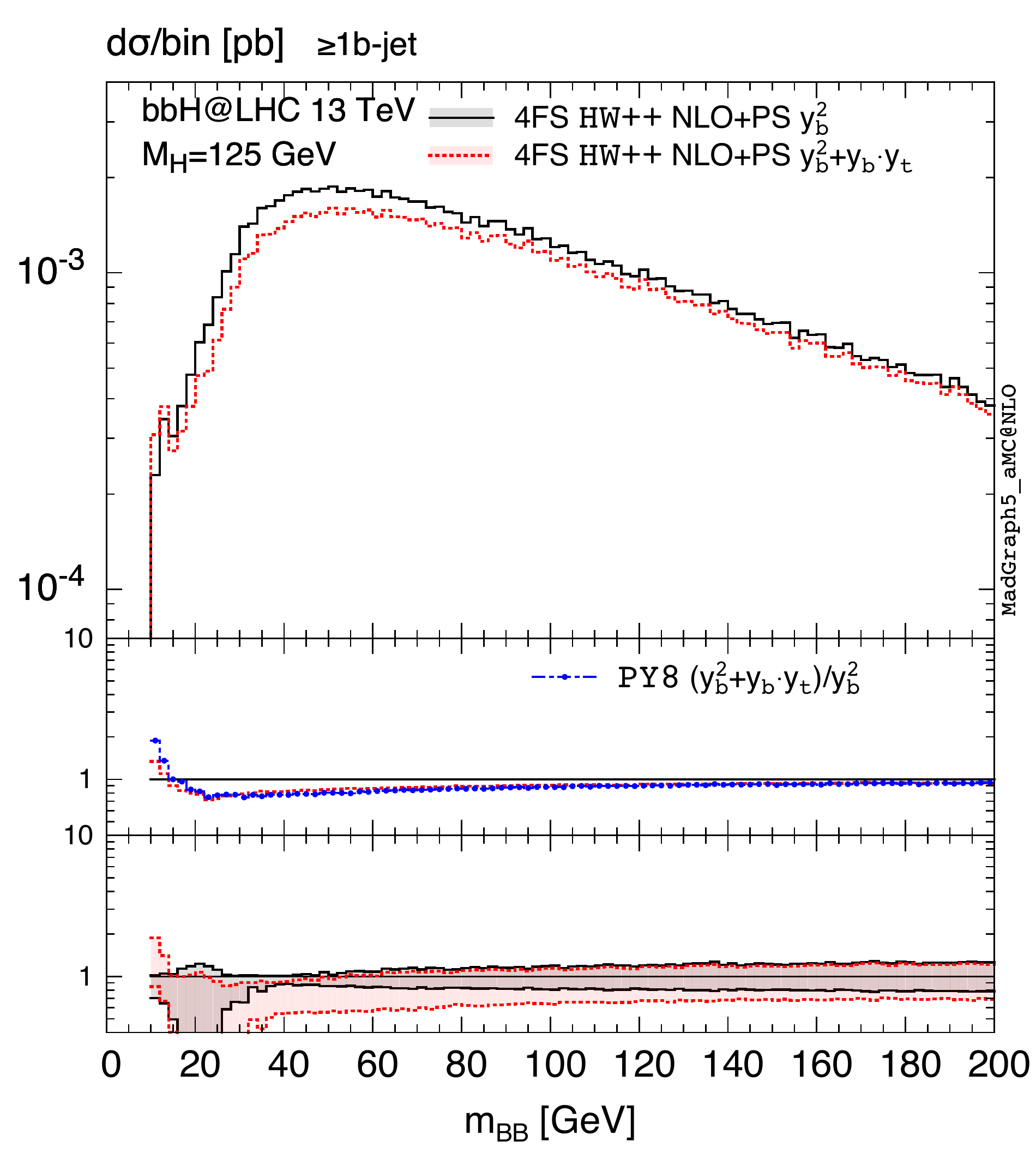,width=0.48\textwidth}
\caption{\label{fig:yb2vsybyt1} 
Invariant mass of the two hardest $B$ hadrons, with or without the
$\yb\yt$ contribution. The left panel presents the fully-inclusive
case, while in the right panel the presence of at least one $b$-jet 
is required. See the text for details.
}
\end{center}
\end{figure}
The results for $m_{BB}$ are displayed in fig.~\ref{fig:yb2vsybyt1},
at the NLO+PS accuracy with \HWpp. The black solid histogram is the
result for $\sigma_{\yb^2}$, and the red dotted histogram the result
for \mbox{$\sigma_{\yb^2}+\sigma_{\yb\yt}$}. The left panel presents
the inclusive case, while in the right panel we have required the
presence of at least one $b$-jet. We start by commenting the inclusive
case: as one can see, the two predictions are 
in very good agreement at large $m_{BB}$ (up
to a rescaling), and in violent disagreement close to threshold, where
the striking feature is a sharp peak which originates from the 
$\yb\yt$ contribution. Given the situation of fig.~\ref{fig:4FSmBB},
and in particular the large differences between \HWpp\ and \PYe\ there, 
in the upper inset we report the ratio
\mbox{$(\sigma_{\yb^2}+\sigma_{\yb\yt})/\sigma_{\yb^2}$} not only
for \HWpp\ (i.e.~the ratio of the two curves displayed in the main 
frame; red dotted), but also for \PYe\ (blue dot-dashed). 
As one can see, the two ratios are in rather
good agreement with each other; in other words, the pattern of 
the peak vs no-peak structure close to the threshold 
of $m_{BB}$ is essentially MC-independent, and is thus
purely of matrix-element origin. It is in fact straightforward to connect
this behaviour to the topology of the one-loop graphs that contribute 
to the $\yb\yt$ term (see for example fig.~\ref{fig:4FSoneloop:a}, 
\ref{fig:4FSoneloop:b}, and~\ref{fig:4FSoneloop:e}): the vast majority
of them feature a $g\to b\bb$ splitting, which is naturally enhanced
at small $m_{BB}$ (in the case of the $\yb^2$ contribution, only a small
fraction of diagrams contain such a splitting). The enhancement due to the
$g\to b\bb$ splitting is easily countered, for example by requiring
the presence in the event of a $b$-jet: the result is shown in the
right panel of fig.~\ref{fig:yb2vsybyt1}, where the sharp peak at
small $m_{BB}$ does not appear any longer; the effect of the 
$\yb\yt$ terms is now much milder (and the relative behaviour of 
\HWpp\ and \PYe\ is again almost identical). We have verified that,
in a completely analogous manner, the $\yb\yt$ contribution is quite
flat also in the case of the invariant mass of the two hardest $b$-jets.

\begin{figure}[t]
  \begin{center}
    \epsfig{figure=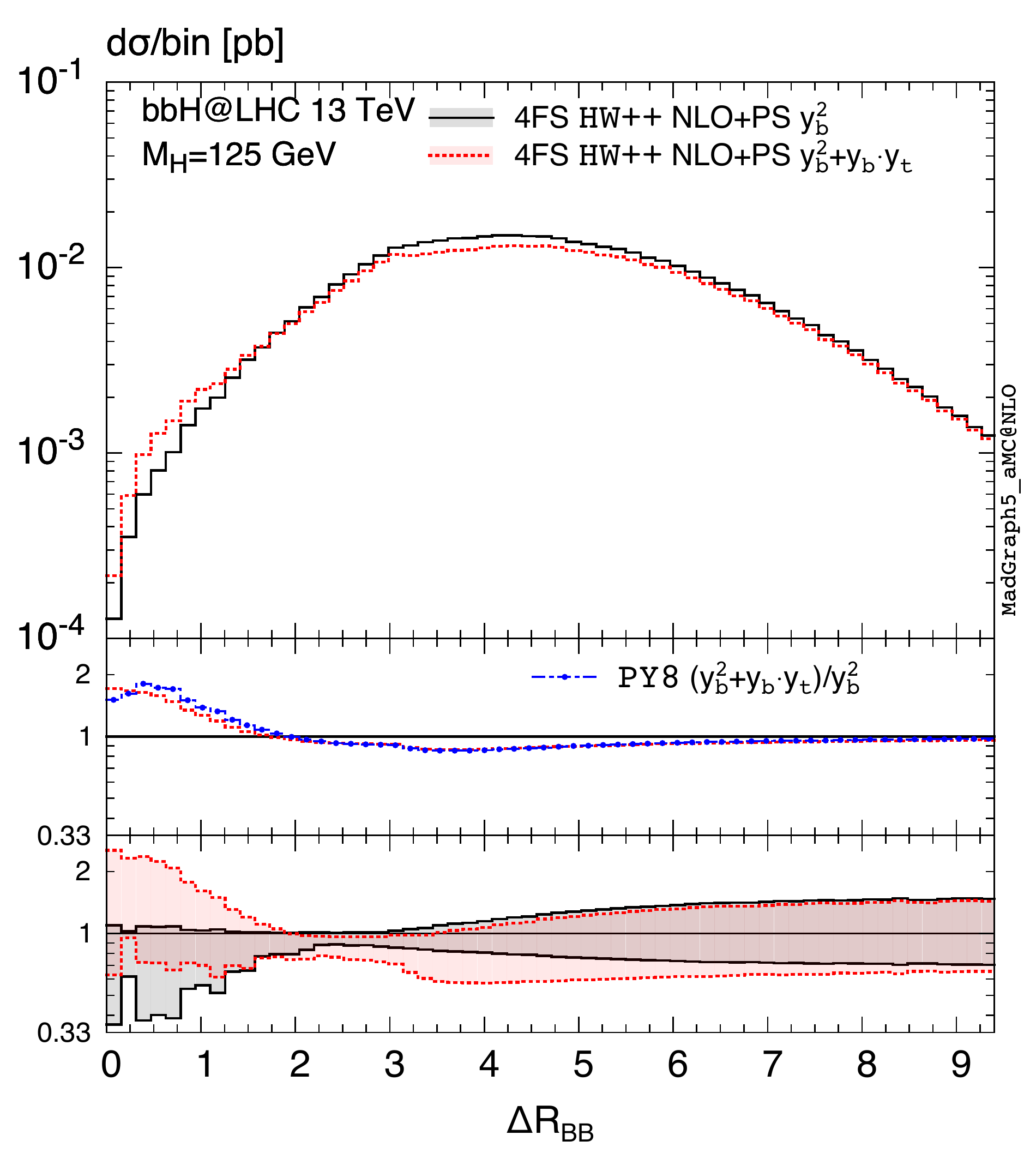,width=0.48\textwidth}
    \epsfig{figure=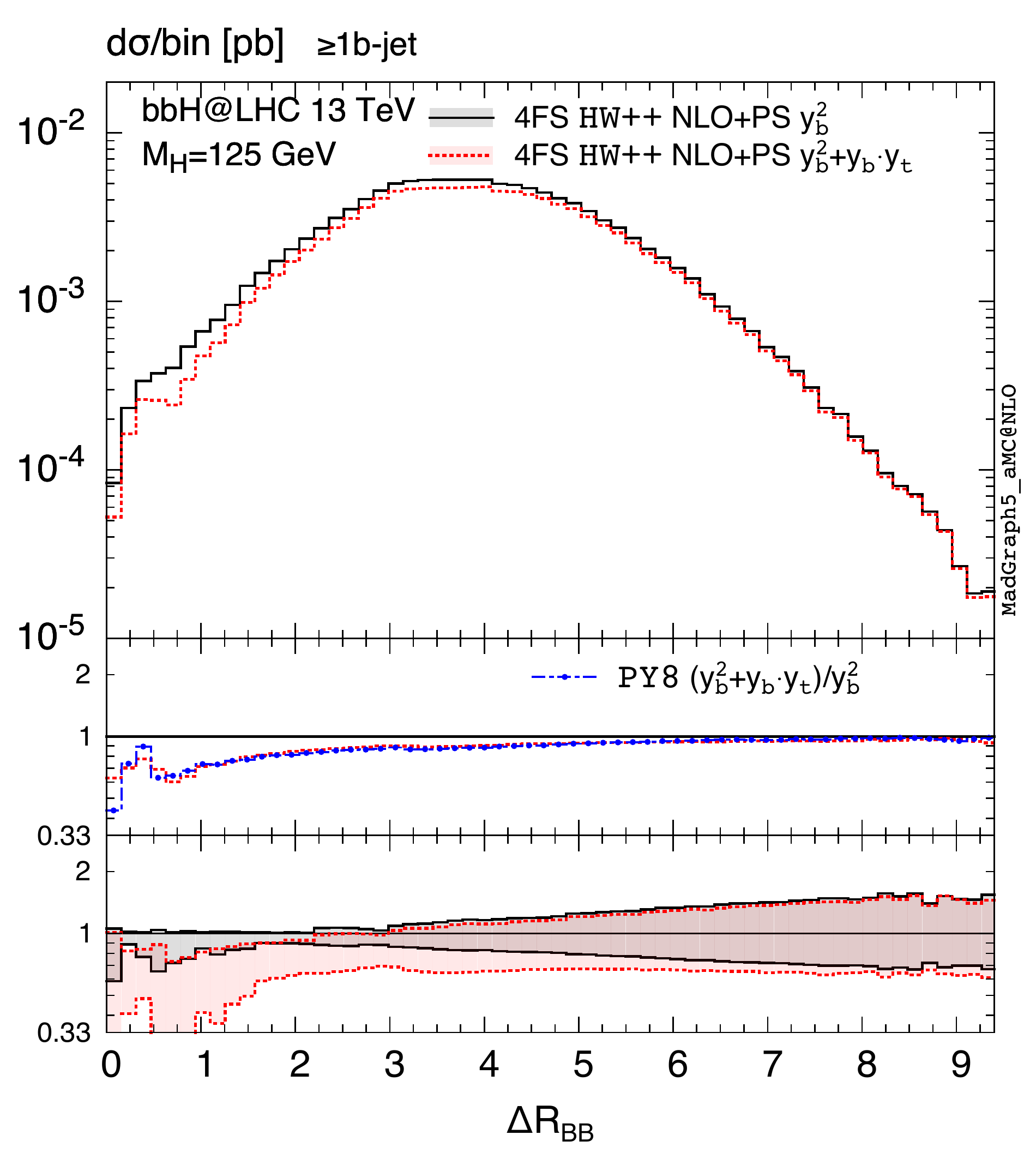,width=0.48\textwidth}
\caption{\label{fig:yb2vsybyt2} 
Same as in fig.~\ref{fig:yb2vsybyt1}, for the distance 
in the $\eta\!-\!\varphi$ plane.
}
\end{center}
\end{figure}
The case of $\Delta R_{BB}$ is presented in fig.~\ref{fig:yb2vsybyt2},
which has the same layout as fig.~\ref{fig:yb2vsybyt1}. The impact
of the $\yb\yt$ contribution is not as outstanding for the present
observable as for $m_{BB}$, but it is still clearly visible,
in that it causes the \mbox{$\sigma_{\yb^2}+\sigma_{\yb\yt}$} cross
section to be larger than the $\sigma_{\yb^2}$ one in the small 
$\Delta R_{BB}$ region, in the inclusive case (see the left
panel of fig.~\ref{fig:yb2vsybyt2}). As already for $m_{BB}$, the
requirement that there be at least one $b$-jet changes the picture
(see the right panel of fig.~\ref{fig:yb2vsybyt2}); still, the $\yb\yt$ 
contribution is less flat than for the vast majority of the other observables, 
and is at the border of the $\sigma_{\yb^2}$ scale-dependence band.

\subsection{Four- and five-flavor scheme comparison\label{sec:4FSvs5FS}}
In this section we present selected 5FS predictions, and compare
them directly to their 4FS counterparts, many of which have been already
shown in sect.~\ref{sec:4FS}. We mostly work at the NLO+PS level,
although occasionally we shall use LO+PS results as well; furthermore,
where appropriate we shall also compare our predictions to those
of (N)NLO+(N)NLL analytical resummations~\cite{Harlander:2014hya}. 
Since the $\yb\yt$ term vanishes in the 5FS up to ${\cal O}(\as^2)$,
we only consider the $\yb^2$ contribution
throughout this section. We also point out that, in the case of
a 5FS computation, one has $s_0=\mH^2$ (see eq.~(\ref{murange})).
Therefore, at variance with the case of the 4FS, $\Qshow$ naturally
assumes values rather smaller than the Higgs mass given the default 
$f_i$ parameters. Thus, in the present case the default choice $\alpha=1$
is perfectly adequate; we have in any case verified that the dependence 
on $\alpha$ is moderate (${\cal O}(10\%)$, and affecting mostly the
intermediate-$\pt$ region), and much smaller than for 4FS results.

\begin{figure}[t]
  \begin{center}
    \epsfig{figure=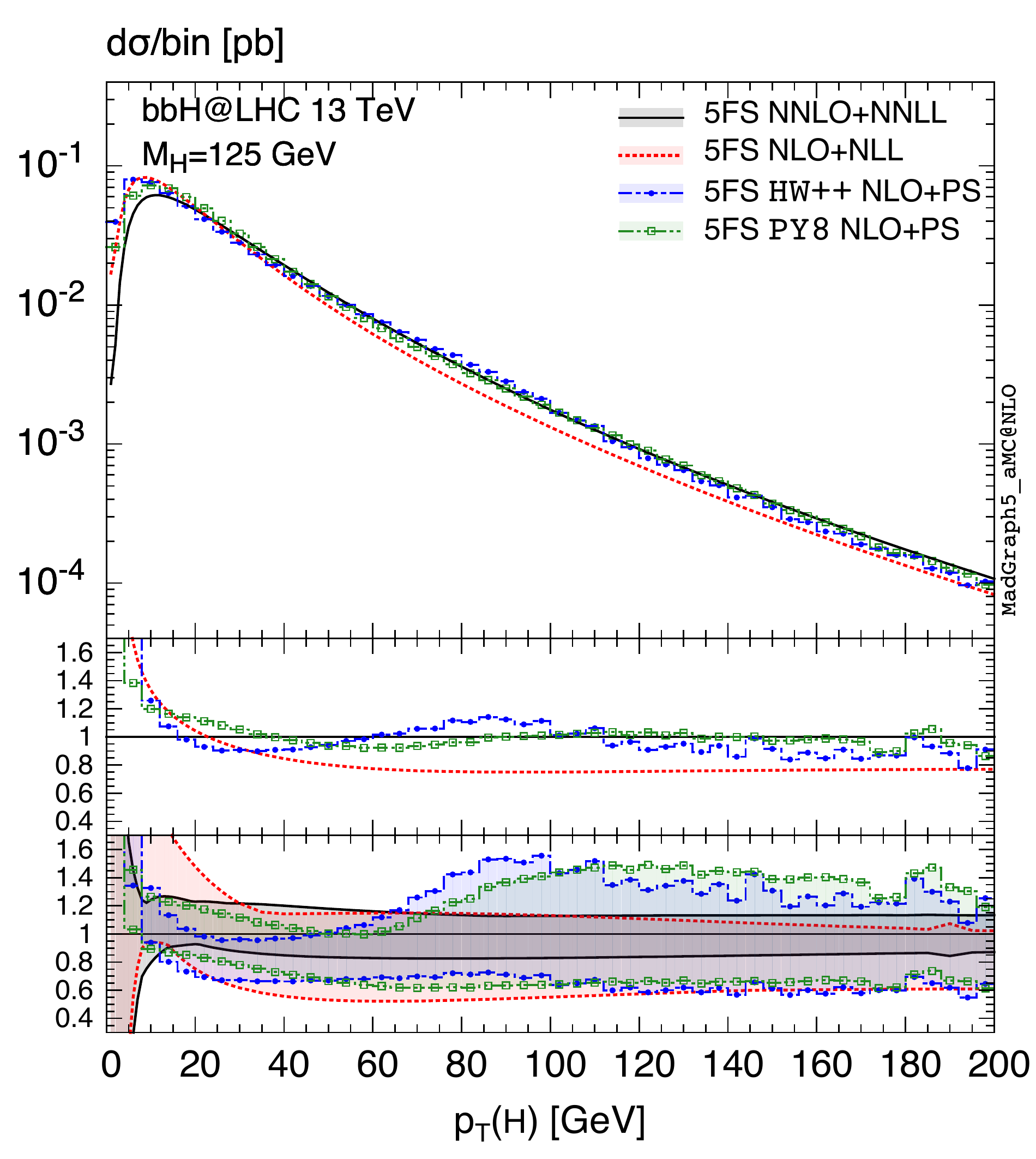,width=0.48\textwidth}
    \epsfig{figure=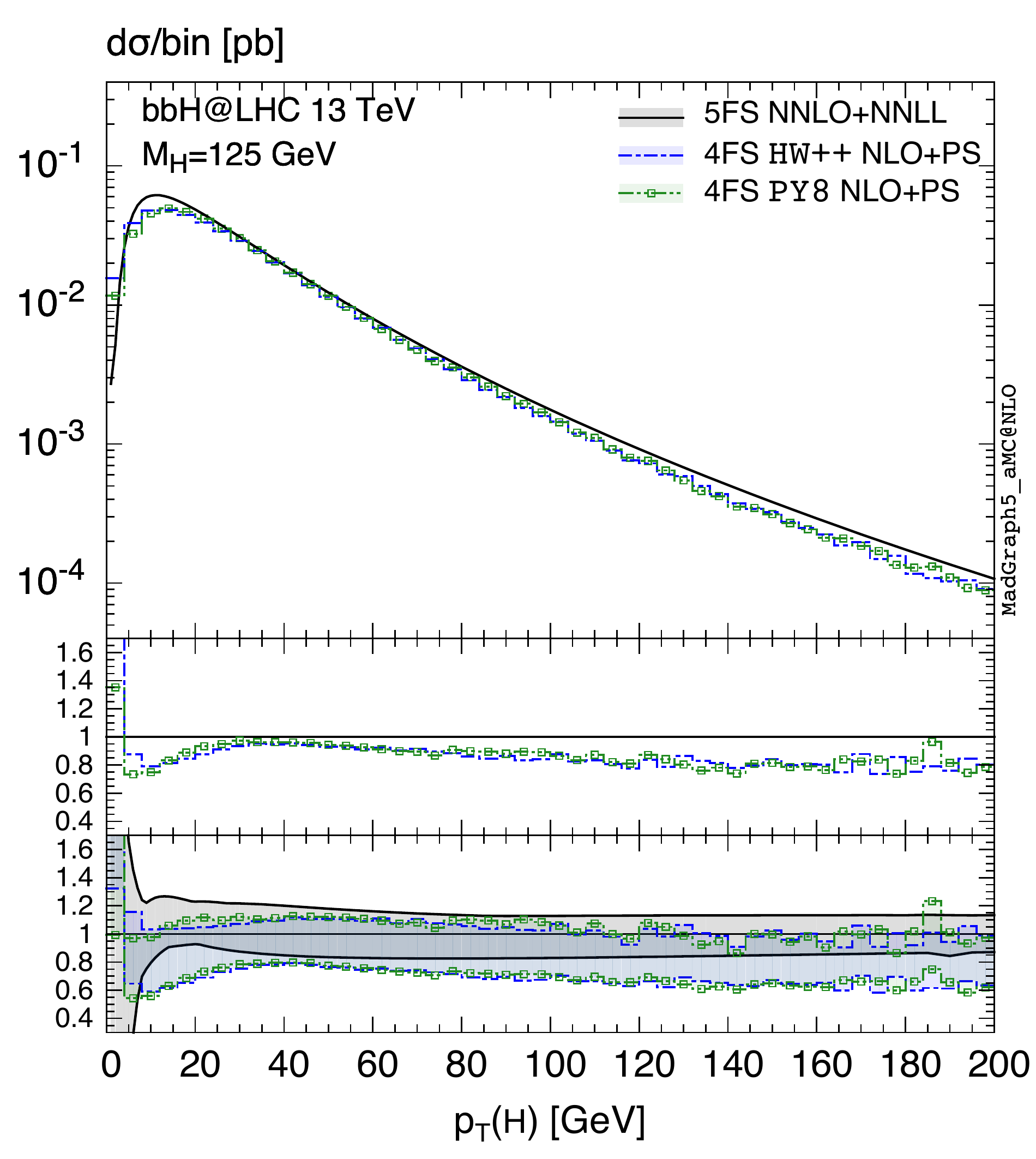,width=0.48\textwidth}
\caption{\label{fig:pTH} 
Higgs transverse momentum. In the left panel analytically-resummed and
NLO+PS 5FS results are compared. In the right panel the NNLO+NNLL prediction
is compared to NLO+PS 4FS results.
}
\end{center}
\end{figure}
\begin{figure}[t]
  \begin{center}
    \epsfig{figure=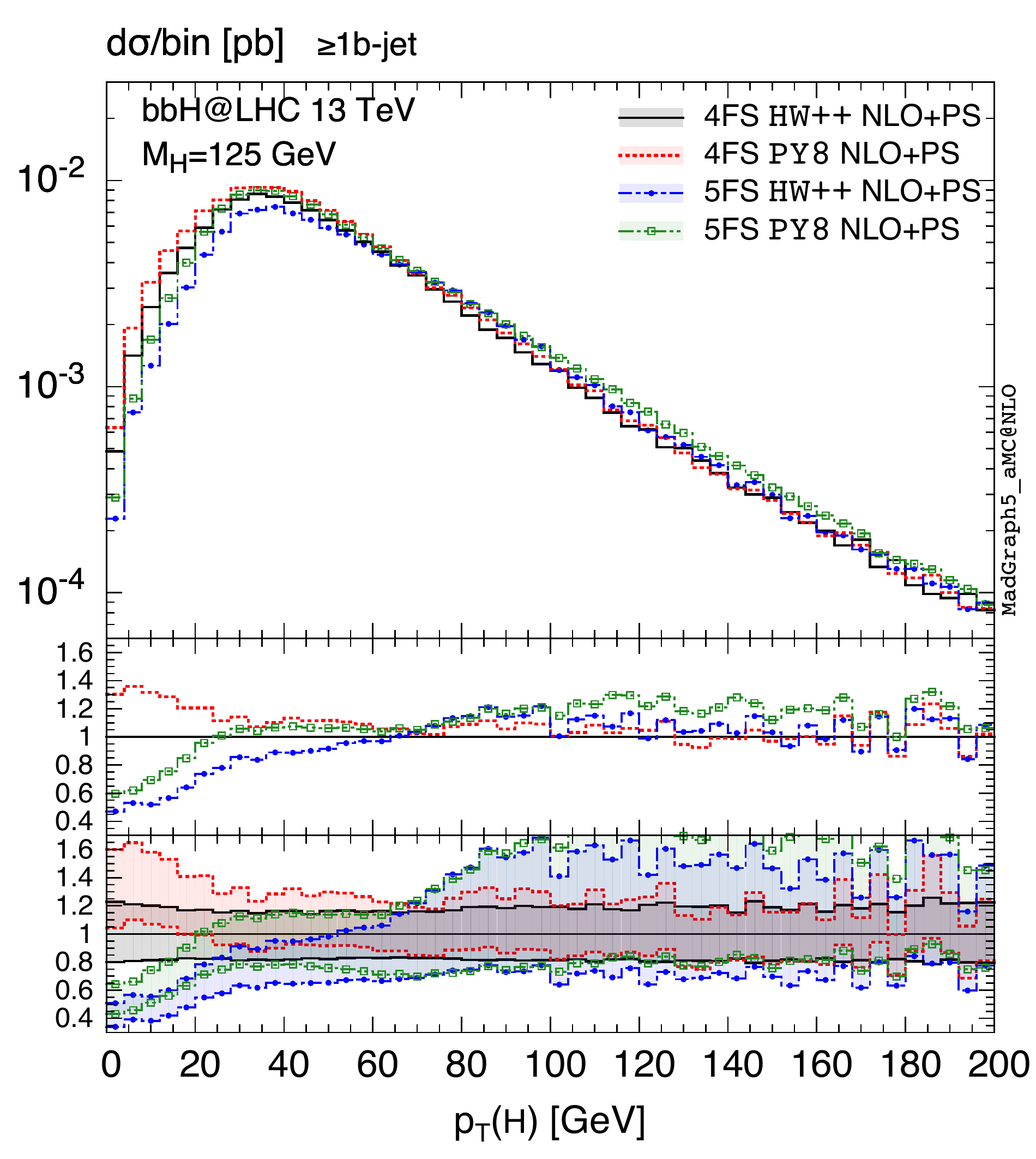,width=0.7\textwidth}
\caption{\label{fig:pTH1bj} 
Higgs transverse momentum in the presence of at least one $b$-jet;
we compare NLO+PS 4FS and 5FS predictions.
}
  \end{center}
\end{figure}
We start by considering the Higgs transverse momentum. In the left
panel of fig.~\ref{fig:pTH} the NLO+PS \HWpp\ and \PYe\ 5FS predictions
(blue dot-dashed and green dash-double-dotted respectively)
are compared with the results of the analytical resummation 
at the NLO+NLL (red dotted) and NNLO+NNLL (black solid) accuracy; the
latter result is the reference curve, used as the denominator in the
computation of the ratios which appear in the insets. The analytically
resummed results, in this plot and in all those which will follow, have been 
rescaled by the bin size for a direct comparison of {\em both rate and shape} 
with the NLO+PS predictions (since the latter are always in the form 
of cross sections per bin, while the former are returned by the code of
ref.~\cite{Harlander:2014hya} as differential distributions); their
associated uncertainty bands (which in the case of the NNLO+NNLL result
are smaller than those relevant to any other simulation considered here) 
include an independent variation of $\Qres$.
In the small-$\pt$ region, there is a good agreement among the two 
NLO+PS and the NLO+NLL results, while the NNLO+NNLL one is visibly lower. 
The peaks of all curves lie within 5~GeV of each other, that of \HWpp\ 
(NNLO+NNLL) being the lowest (highest). Starting from about 
$\pt(H)\sim 60$~GeV, the NLO+PS results are closer to the NNLO+NNLL
curve than to the NLO+NLL one (which however is within the scale uncertainty
bands of the former), the agreement being particularly good in the case 
of \PYe. Note that at large $\pt$ the NNLO and NLO predictions are quite
close to each other; this is analogous to what has been observed in 
ref.~\cite{Harlander:2011fx} for the transverse momentum of 
the hardest jet, and is a consequence of using $\muR=\muF=m_T(H)/4$.
In the right panel of fig.~\ref{fig:pTH} we compare the most accurate
5FS prediction, namely the analytically-resummed NNLO+NNLL, with the NLO+PS
\HWpp\ and \PYe\ ones in the {\em four-flavour} scheme, which have already
appeared in fig.~\ref{fig:pTH4FS}. As we know from that figure, the
agreement between the two NLO+PS results is excellent; what one sees
from fig.~\ref{fig:pTH} is that these NLO+PS predictions also agree
rather well with the NNLO+NNLL one (and in an excellent manner shape-wise), 
except when very close to $\pt(H)=0$, with fully overlapping uncertainty 
bands. 
In terms of shape, the NLO+PS 5FS results also show a comparable
level of agreement for $\pt\gtrsim 20$~GeV, while being noticeably
worse at lower transverse momentum values.
However, one must bear in
mind that the NLO+PS matching systematics (see eq.~(\ref{murange})) is much 
larger in the 4FS than in the 5FS. Conversely, note that the widths of the 
NLO+PS 5FS uncertainty bands are larger than those relevant to the 4FS for 
$\pt(H)\gtrsim 80$~GeV, because from the perturbative viewpoint that
kinematic region is effectively described at the LO in the 5FS.

The $\pt(H)$ distribution is severely affected by the requirement that 
there be at least one $b$-jet in the final state. In fig.~\ref{fig:pTH1bj} 
we present the relevant results, obtained at the NLO+PS accuracy in
the 4FS (black solid (\HWpp) and red dotted (\PYe)) and in the 5FS
(blue dot-dashed (\HWpp) and green dash-double-dotted (\PYe)).
For $\pt(H)\gtrsim 50$~GeV all predictions are within 30\% of each
other, the agreement among the two 4FS results and the 5FS \HWpp\
one being particularly good. Below 50~GeV more significant deviations
(especially in shape)
\begin{figure}[t]
  \begin{center}
    \epsfig{figure=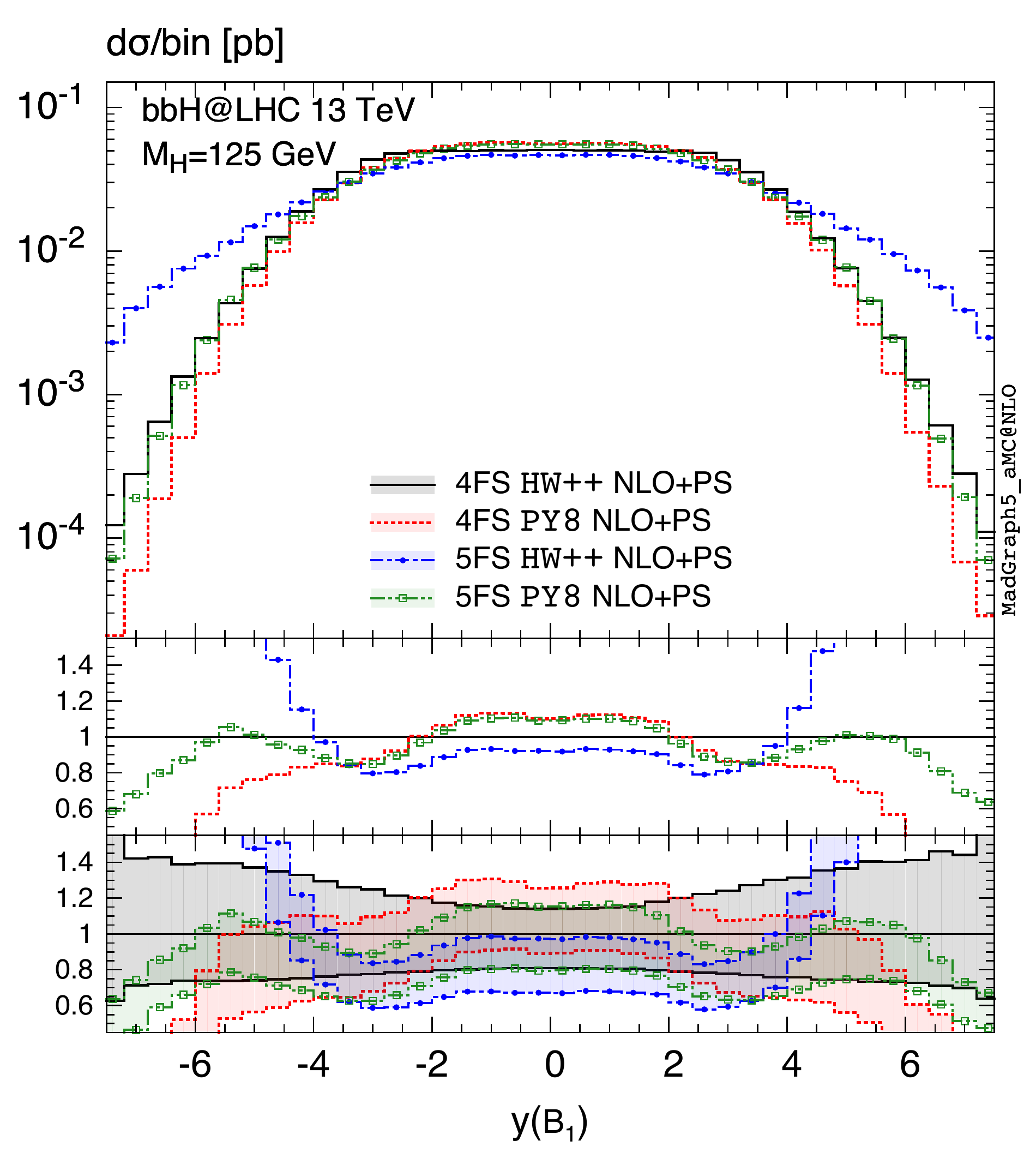,width=0.48\textwidth}
    \epsfig{figure=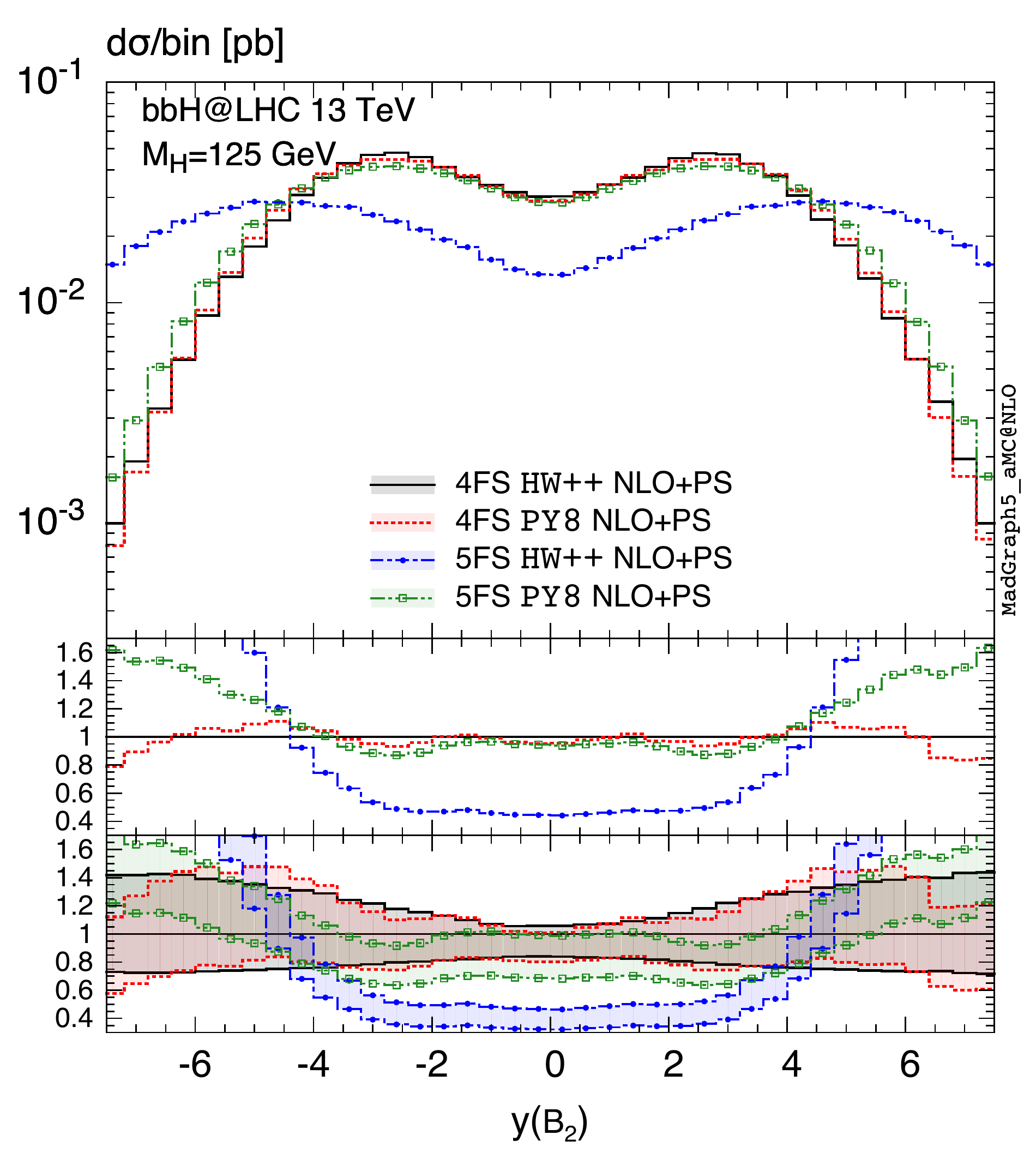,width=0.48\textwidth}
\caption{\label{fig:y1y2all} 
Rapidity of the hardest (left panel) and second-hardest (right panel)
$B$ hadron, in the 4FS and 5FS at the NLO+PS accuracy, as predicted by
\HWpp\ and \PYe. All histograms have been normalised so that their
integrals are equal to one.
}
\end{center}
\end{figure}
start to appear, which increase with decreasing $\pt(H)$. The pattern
of the comparison between the two MCs is the same in the two schemes:
namely, the \PYe\ cross section is larger than the corresponding 
\HWpp\ one. The two 4FS results are larger than the two 5FS ones. 
Below $\pt(H)\sim 10$~GeV, the uncertainty bands of the 4FS results
do not overlap any longer with those of the 5FS ones; within a given scheme, 
the bands do overlap, but the central predictions show differences at the
level of 30\% and 20\% in the 4FS and 5FS respectively. In conclusion,
although the overall agreement between the 4FS and 5FS results is
reasonable, shape-wise visible discrepancies do appear, which would
thus be interesting to investigate using data, especially in view
of the fact that theoretically, for an observable that features a
tagged $b$-jet, the 4FS is expected to be superior to the 5FS.

In the context of MC simulations the presence of a massless $b$ quark
in the initial state at the matrix-element level poses a non-trivial 
problem; apart from the necessity of evolving it backwards in a way 
that matches the flavour content of the incoming hadron, one always faces 
a kinematic constraint, imposed by the fact that eventually the $b$ quark 
will have to appear in the final state, with a mass of around 5~GeV.
This problem, which is essentially process-independent, is particularly
severe in the case of \HWs~\cite{Corcella:2000bw,Corcella:2002jc}, 
where it leads to strangely-looking 
distributions, with longitudinal observables being particularly affected. 
An explicit example is given in fig.~14 of ref.~\cite{Frixione:2010ra},
where 5FS single-top production has been considered; from that figure,
one can also see how this issue has been addressed by \HWpp, which
(almost completely) rectifies the behaviour of \HWs. It is important
to stress that it is the region $\pt\sim 0$ which is relevant here:
as soon as one imposes a {\em realistic} tagging condition (by requiring
a minimum $\pt$ in order to observe a $B$ hadron), or is sufficiently
inclusive, the problem above becomes essentially irrelevant. Still,
from the theoretical point of view this small-$\pt$ behaviour of
$b$-flavoured hadrons is less than satisfactory. 
\begin{figure}[t]
  \begin{center}
    \epsfig{figure=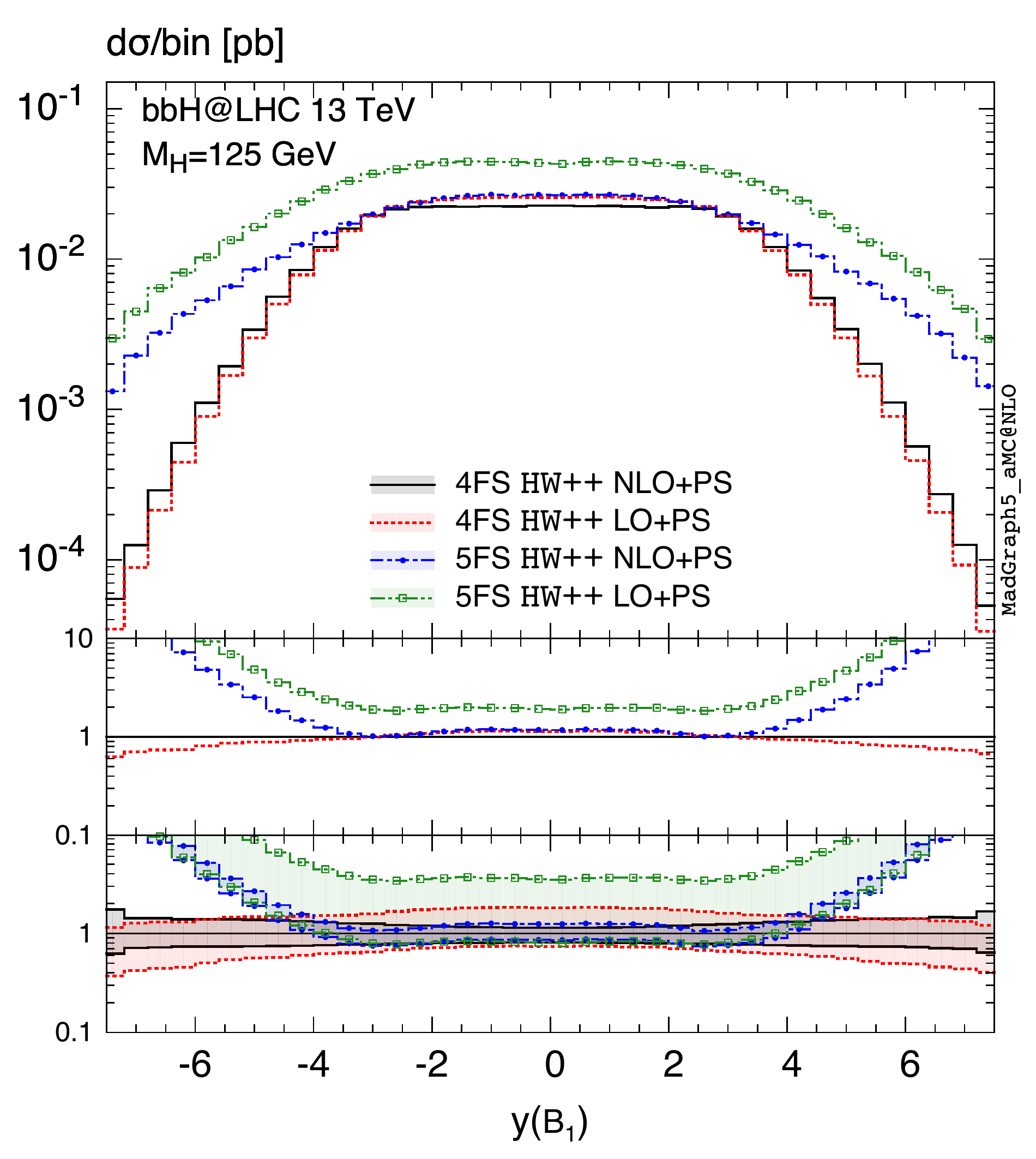,width=0.48\textwidth}
    \epsfig{figure=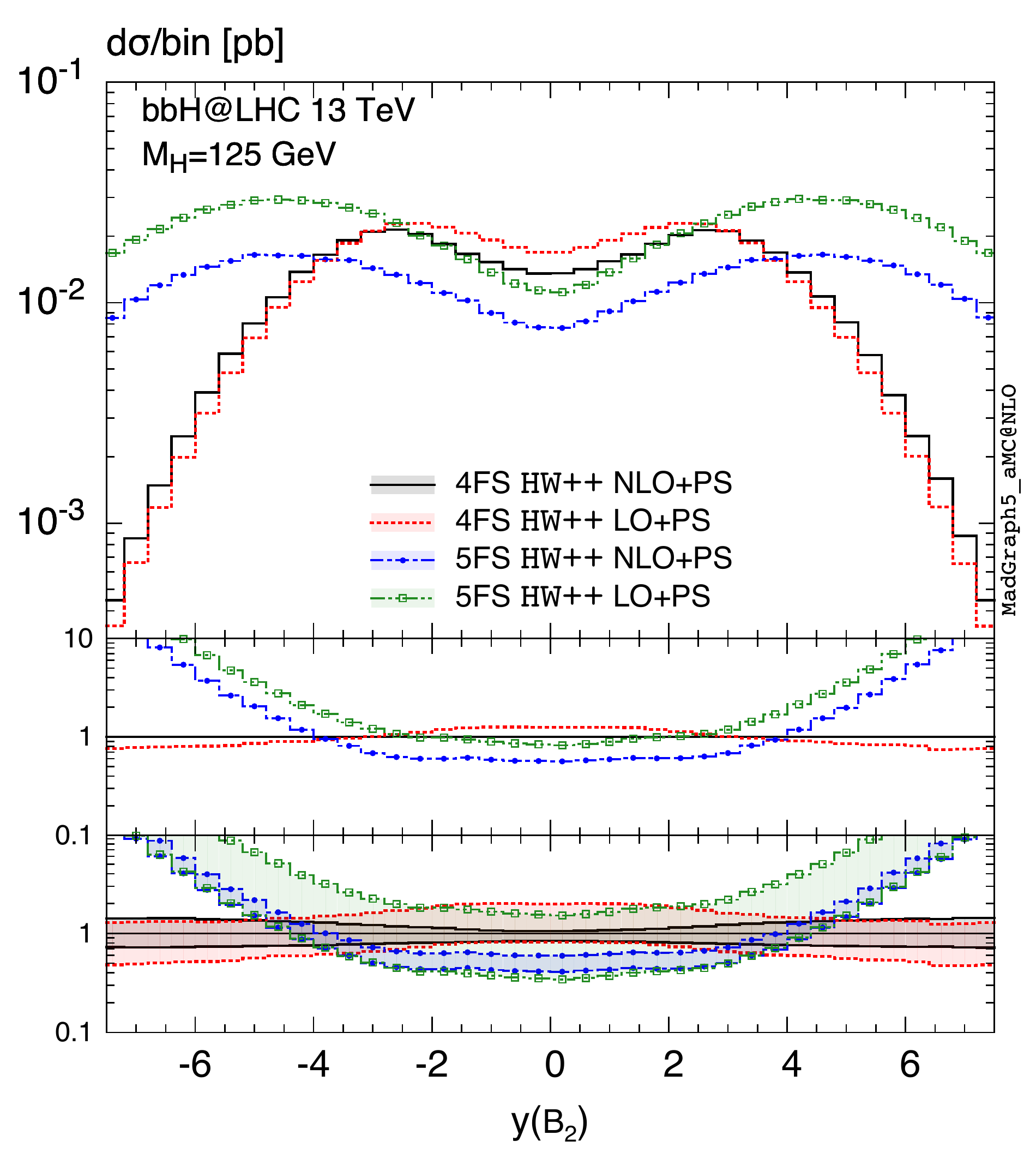,width=0.48\textwidth}
\caption{\label{fig:y1y2hwpp} 
Same as in fig.~\ref{fig:y1y2all}, for \HWpp\ at the NLO+PS and LO+PS
accuracy. The normalisation is absolute here.
}

\end{center}
\end{figure}

With this in mind, and given the improvement of \HWpp\ w.r.t.~\HWs\ in 
the case of single-top production, we have considered the rapidity 
distributions of the hardest and second-hardest $B$ hadrons. The 
(normalised) results are displayed in the left and right panel, respectively, 
of fig.~\ref{fig:y1y2all}, where we show both the 4FS predictions
(black solid (\HWpp) and red dotted (\PYe)) and the 5FS ones
(blue dot-dashed (\HWpp) and green dash-double-dotted (\PYe)).
The prominent feature of these plots is the behaviour of the
\HWpp\ 5FS results, which are vastly broader than all of the other
three curves. Therefore, although no features appear such as those 
mentioned above for \HWs, it is likely that \HWpp\ still tends to
produce $B$ hadrons too close to the beam line when simulations
are performed in the 5FS. As far as the other histograms are concerned,
the agreement between the two \PYe\ predictions is extremely good
for rapidities as large as about $3$. Beyond this value, the 5FS result
is broader than the 4FS one, and is actually rather close to the
\HWpp\ 4FS result in the case of the hardest $B$ hadron (not so
for the second-hardest $B$ hadron, for which at large rapidities
the two 4FS predictions agree at the level of 20\%). 

In order to further the study of the behaviour of \HWpp\ for these
rapidities, in fig.~\ref{fig:y1y2hwpp} we present the comparison between
the NLO+PS and LO+PS predictions, in both the 4FS and 5FS. As one can
see, the effect of the NLO corrections is fairly modest: the 5FS
NLO results are slightly more central than their LO counterparts,
and the opposite happens in the 4FS. This implies that the two NLO+PS
predictions are marginally closer to each other than the two LO+PS ones,
but this fact is quite irrelevant given the vastly different shapes
one obtains in the two schemes.

\begin{figure}[t]
  \begin{center}
    \epsfig{figure=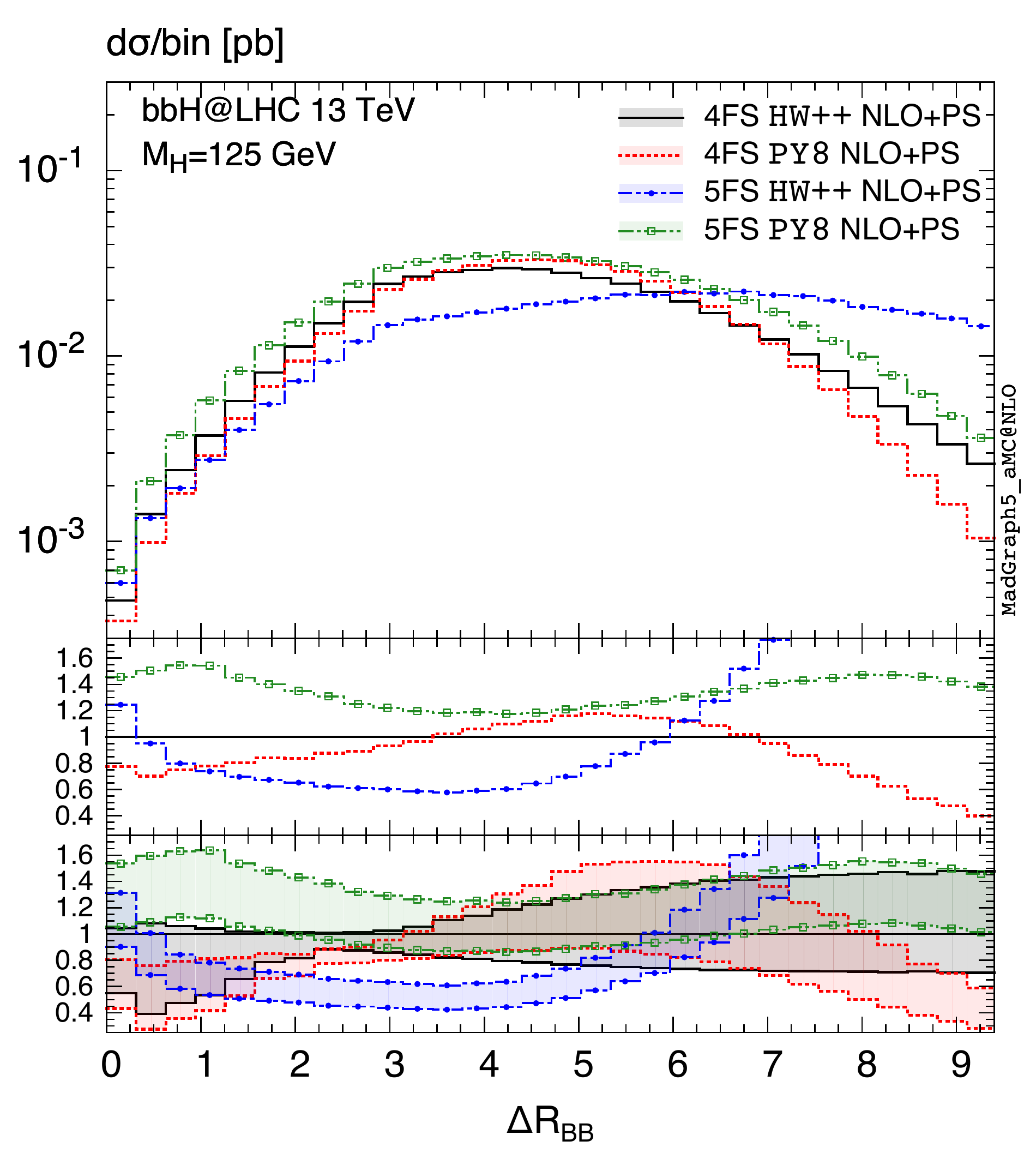,width=0.48\textwidth}
    \epsfig{figure=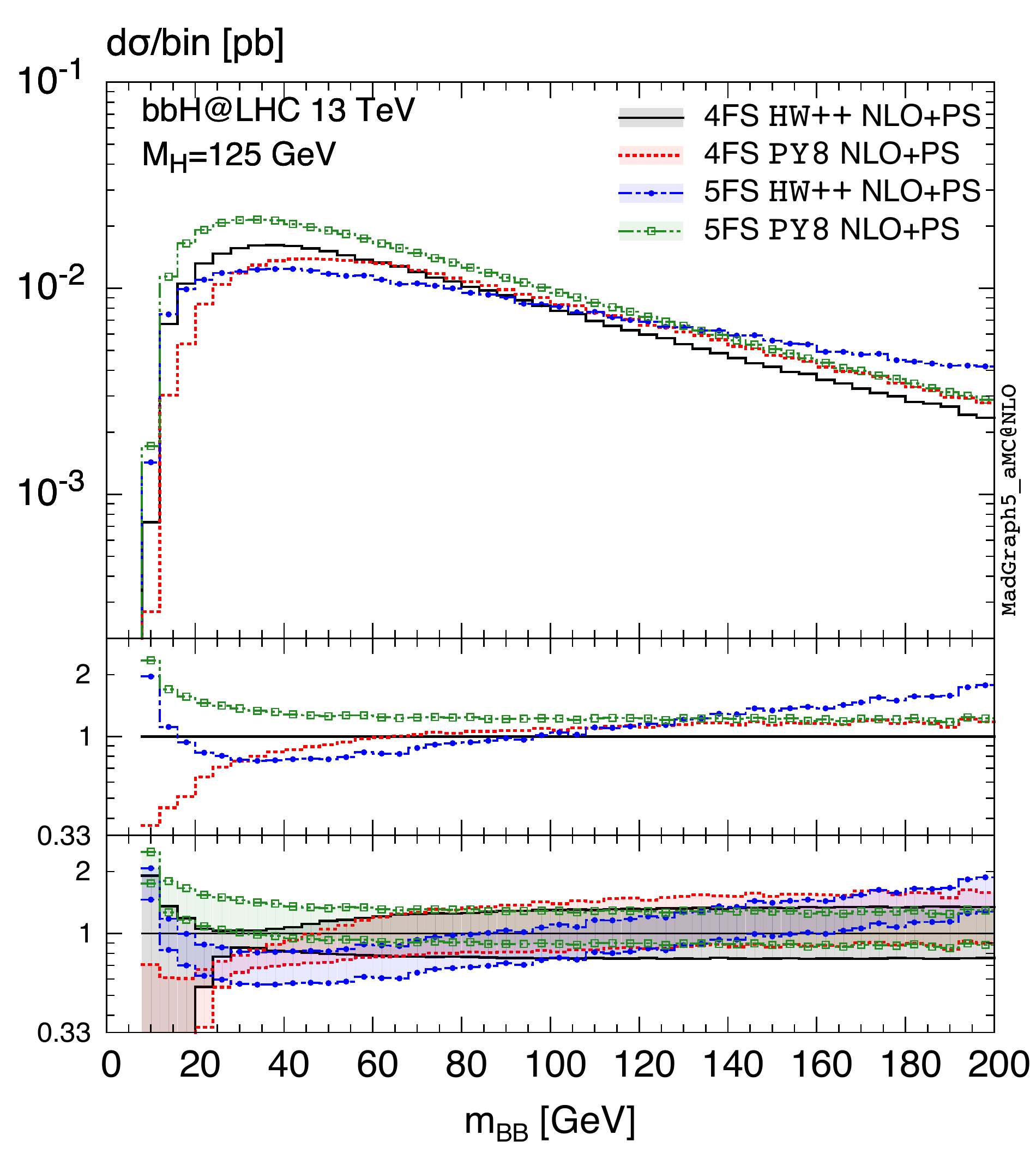,width=0.48\textwidth}
\caption{\label{fig:mBBall}
Comparison between the 4FS and 5FS results for the invariant mass of
(left panel) and the $\eta\!-\!\varphi$ distance between (right panel) the
two hardest $B$ hadrons.
}

\end{center}
\end{figure}
The observations above have bearings on the predictions for the
two quantities which have been already addressed at length in
sect.~\ref{sec:4FS}, namely the invariant mass of ($m_{BB}$), and 
the $\eta\!-\!\varphi$ distance between ($\Delta R_{BB}$), the
two hardest $B$ hadrons. We display these observables in
fig.~\ref{fig:mBBall}, by comparing the 4FS predictions already
presented in fig.~\ref{fig:4FSmBB} and fig.~\ref{fig:4FSDeltaRBB} 
to their 5FS counterparts. The most evident feature in the two
panels of fig.~\ref{fig:mBBall} is the very large \HWpp\ 5FS cross
section at large $m_{BB}$ and $\Delta R_{BB}$, which suggests again
that \HWpp\ in this scheme tends to produce $B$ hadrons fairly close
to the beam line, and in opposite directions. The other three results
are much closer to each other, although large differences remain; one
may note how the 4FS \HWpp\ predictions have shapes rather similar to
the 5FS \PYe\ ones. There is no fundamental reason why this should be
so; in particular, one should bear in mind that for these observables
one expects the 4FS to be significantly more reliable than the 5FS.
Still, our results indicate that the underlying MCs, even in the context
of matched simulations at the NLO, play a non-negligible role.
A thorough comparison with data, for this and other $b$-initiated
processes, will certainly be beneficial for a better understanding
of these issues, and for improving the tuning of the relevant
long-distance parameters.

\section{Conclusions\label{sec:conclusions}}
In this paper we have studied the hadroproduction of a Higgs
boson in association with $b$ quarks, and presented for 
the first time NLO QCD predictions matched to parton showers.
We have worked in the \aNLO\ framework, endowed with the capability,
which was unavailable in the code prior to this study, of treating the 
renormalisation of the bottom Yukawa coupling in a fully flexible manner,
and in particular in the $\msbar{}$ scheme.
We have provided results in both the four-flavour and the five-flavour
schemes, which we have compared to each other and, in the case of the
Higgs transverse momentum, with analytically-resummed (N)NLO+(N)NLL results
as well. We have also considered the ${\cal O}(\yb\yt\as^3)$ term in the 4FS, 
which might be viewed as the leading contribution to the interference between 
the $\bbH$ and gluon-fusion channels.

The key feature of the predictions we report is their being fully
differential, independently of whether they are matched with
parton showers. We have documented this by discussing the cases of several 
observables that are exclusive in the Higgs and/or in the $b$-flavoured 
hadron or $b$-quark momenta, and which are notable for their 
characteristics. Although such observables represent only a limited 
sample of what can be obtained by means of \aNLO, they extend in a 
very significant manner the knowledge of differential quantities which 
was available in the literature so far. One aspect of our results that 
is particularly worth stressing is that their associated scale (and PDF,
which for simplicity we have refrained from investigating here)
uncertainties can be computed at the same time as the central results
without any noticeable CPU cost. This is particularly important in
the case of $\bbH$ production, in view of the large theoretical systematics 
which affect this process, and which must thus be carefully taken into 
account, as we have done for all predictions given in sect.~\ref{sec:results}.

Because of the fully-exclusive nature of our computations, we
assume the 4FS results, especially thanks to the matching to parton
showers, to be generally superior to the 5FS ones, and we believe that
they should constitute the default choice for any realistic physics 
simulation. Having said that, for observables that are fully inclusive
in the degrees of freedom of the associated $b$ quarks, or for which
$\mb$ is negligible (e.g.~at large $\pt$'s) the differences between
the four- and five-flavour schemes must be carefully assessed.
From the phenomenology viewpoint, the main conclusions of 
this work are the following.
\begin{itemize}
\item We have found evidence that relatively small values for the 
shower scales in 4FS computations have to be preferred. This is in
keeping with the by-now standard choices for the other hard scales that
enter the process, and with analogous findings in the context of other 
types of calculations (i.e.~fixed order in both the four- and five-flavour
schemes, and analytical resummations).
\item The impact of NLO QCD corrections is very significant, both in
terms of reducing the scale uncertainties w.r.t.~those that affect the
LO results, and in the changes they induce in the shapes of {\em some}
differential observables (in particular, the $m_{BB}$ and $\Delta R_{BB}$
correlations). Even at the NLO, however, the perturbative 
systematics that affect $\bbH$ production are sizable.
\item The ${\cal O}(\yb\yt\as^3)$ interference contribution reduces the 
inclusive rate by about 10\%. It is generally flat in the phase space, 
except close to the threshold of the invariant mass of the hardest-$B$-hadron 
pair, where it gives a very prominent peak structure. Such a peak tends
to disappear when increasing the minimum transverse momentum of the
$B$ hadrons, or when $b$ jets are employed.
\item The matching of NLO results with parton showers plays a
very important role. There are observables for which 
fixed-order predictions are significantly different w.r.t.~those after 
showers. On the other hand, there are cases where the \PYe\ and \HWpp\
results show large discrepancies, owing to fundamental differences
in the implementation of core physics characteristics (such as 
shower and hadronisation mechanisms); this is especially true for
5FS simulations. The MC systematics must thus be considered very 
carefully in an observable-by-observable manner.
\item With the exception of the rapidity distributions of $b$-flavoured 
hadrons as predicted by \HWpp{}, the agreement at NLO+PS between 4FS and
5FS results is generally good, and particularly so in the case of \PY8.
For observables exclusive in the associated $b$ hadrons this is perhaps
surprising, and appears to suggest that the underlying parton shower
description of branchings that involve massless $b$ quarks is adequate
in most cases.
\item The NLO+PS predictions for the Higgs transverse momentum, inclusive
in the degrees of freedom of the $b$ quarks, compare remarkably well with 
the analytically-resummed ones. In terms of shape, this is particularly
true for the 4FS and the NNLO+NNLL results.
\end{itemize}

\section*{Acknowledgements}
We are grateful to Marco Zaro for his help and collaboration at various 
stages. This work has been performed in the framework of the ERC grant 291377 
``LHCtheory: Theoretical predictions and analyses of LHC physics: 
advancing the precision frontier", and supported in part
by the Research Executive Agency (REA) of the European Union under 
Grant Agreement number PITN-GA-2012-315877 (MCNet). 
The work of FM is supported by the IISN ``MadGraph'' convention 4.4511.10.
The work of MW is supported by the European Commission through the FP7 
Marie Curie Initial Training Network ``LHCPhenoNet'' (PITN-GA-2010-264564).
The work of PT is supported in part by the Swiss National Science Foundation
(SNF) under contract 200020-149517 and in part by the LHCPhenoNet.
The work of VH is supported by the SNF with grant PBELP2\_146525.

\bibliographystyle{UTPstyle}
\bibliography{bbH_bib}

\end{document}